# Do Pension Benefits Accelerate Cognitive Decline? Evidence from Rural China*

Plamen Nikolov[bcd☆]     Md Shahadath Hossain[a]

**Abstract**

Economists have mainly focused on human capital accumulation, rather than on the causes and consequences of human capital depreciation in late adulthood. To investigate how human capital depreciates over the life cycle, we examine how a newly introduced pension program, the National Rural Pension Scheme, affects cognitive performance in rural China. We find significant adverse effects of access to pension benefits on cognitive functioning among the elderly. We detect the most substantial impact of the program on delayed recall, a cognition measure linked to the onset of dementia. In terms of mechanisms, we find that cognitive deterioration in late adulthood is mediated by a substantial reduction in social engagement, volunteering, and activities fostering mental acuity.

---

*We thank Chris "Kitt" Carpenter, Ted Miguel, Julie Cullen, David Cutler, Jonathan Kolstad, David Canning, Eric Edmonds, Daniel Millimet, Carly Urban, Paul Novosad, Matt Harris, Zoe McLaren, Laura Zimmerman, Susan Godlonton, Carly Trachtman, Brett Matsumoto, Darren Kauffman, Dan Lee, Shelly Lundberg, Livia Montana, Matthew Bonci, Steve Yeh, Evan Riehl, Petra Moser, James MacKinnon, and Morten Nielsen for constructive feedback and helpful comments. Alan Adelman provided invaluable assistance in earlier stages of this project. All errors are our own.

☆Corresponding Author: Plamen Nikolov. Email: pnikolov@post.harvard.edu

[a]State University of New York (Binghamton)
[b]IZA Institute of Labor Economics
[c]Harvard Institute for Quantitative Social Science
[d]Global Labor Organization

# I. Introduction

Across the globe, population aging continues at a rapid pace, and this trend has motivated many developing countries to introduce pension programs. However, the introduction of such programs could have a fortuitous consequence related to cognitive function in old age. There is growing evidence that cognitive activity is linked to enhanced cognitive function (Bonsang et al. 2012), which suggests that long-term involvement in the workforce may be preventative of cognitive aging. Understanding how retirement plans affect cognitive function in old age is crucial for fully recognizing their welfare implications and gaining insight into how cognitive abilities evolve over the life course.

Cognitive abilities represent one dimension of human capital, as do education, health, and noncognitive skills. Historically, the economics literature has primarily focused on human capital formation (Heckman 2000) and considerably less on the causes and consequences of human capital depreciation, including cognitive decline. However, recent neuropsychological evidence suggests that the adult brain is malleable and open to enhancement even in late adulthood (Howard-Jones, Washbrook, and Meadows 2012). Cognitive aging is a complex phenomenon, and its economic and policy causes are not well understood. In this paper, we analyze the effects of a pension program on cognitive performance in old age.

China introduced the New Rural Pension Scheme (NRPS) in 2009 to ease demographic pressures and alleviate concerns about old-age poverty (Dorfman et al. 2013). This paper examines the causal effect of participation in the NRPS on human capital depreciation among individuals aged 60 and over. The NRPS is a contributions-based program; workers who opt into the program make annual contributions matched by government subsidies. Once beneficiaries reach age 60, they can start receiving their annuity benefits irrespective of their retirement status. The annuity comprises a basic pension from the government and a part determined from prior contributions. We analyze the effects of the NRPS on two outcomes: episodic memory and mental intactness. We measure episodic memory through verbal learning and recall tasks (encompassing fluid intelligence). Mental intactness, elicited through a series of memory tasks, generally gauges fluid and crystallized intelligence.

Studying how human capital depreciates over the life cycle has powerful economic consequences. At the micro-level, cognitive functioning is crucial for decision-making. Elderly individuals make complex financial, health, and long-term care decisions, with significant economic implications (Korniotis and Kumar 2011; De Bruin 2017). Given the lack of intermediary market institutions in rural areas to aid with financial decisions connected to income security or health care provision, examining the impact on the cognition of the elderly population in a country like China may be especially crucial. Understanding the causes of cognitive decline is also crucial for policy, as the relationship between cognitive aging and productivity affects long-term economic growth (Meisenberg, 2014).

Our empirical strategy draws on the staggered implementation of the NRPS program across rural parts of China between 2009 and 2013. We use a difference-in-difference-in-differences design (from now on, *DDD estimator*) to identify the causal effect of access to the NRPS on cognition among aging adults. We recover treatment effects based on identifying variation from three distinct sources: the program rollout at the residential community level[1], the timing of the implementation by each municipality, and the fact that the annuity benefit was extended only to participants 60 and over. Our analysis relies on a new data source—the Chinese Health and Retirement Longitudinal Survey (CHARLS)— nationally representative of individuals aged 45 and above.

[1] Our primary data source, the CHARLS, refers to this administrative unit, *shequ*, as *community* (/社区). Hence, from now on we refer to these units as *communities* (or *municipalities*). We caveat that municipalities and *shequs* are not the same administrative entities because municipalities typically have the power to exercise self-government.

Retirement plans typically offer benefits that guarantee participants a certain level of income security in old age (Cutler and Johnson 2004).[2] Nevertheless, we find clear evidence of adverse effects on cognitive performance among NRPS participants. Specifically, we find that the provision of pension benefits negatively impacts immediate recall, delayed recall, and total word recall for program participants. This finding is significant, as lower performance on delayed recall memory measures has been a highly accurate detector of dementia among senior individuals (Welsh et al. 1991). For the comprehensive cognitive performance index, relative to performance on the test before accessing program benefits, the estimated decline is 12 percent of a standard deviation (or approximately five percent of the average baseline score on the measure). This decline occurs about four years after the onset of program benefits. When we benchmark our estimate to general ability measures, we show that a 5-percent drop in the average total word recall score is equivalent to a decline in general intelligence by 1.7 percent (relative to the general population).[3] Our findings are robust to specifications using alternative NRPS participation measures. Furthermore, we find evidence to suggest that prolonged exposure to the program exacerbates cognitive decline.

[2] Feldstein and Liebman (2002) overview the introduction and objectives of social insurance programs in high-income countries.

[3] Using a meta-analysis approach, Ackerman et al. (2005) investigate the relationship between recall memory measures and general intelligence (fluid intelligence). The study finds that a 1-percent decline in word scores leads to a 0.33-percent decrease in proxies of general intelligence.

To better understand the cognitive decline process, we attempt to shed light on the basic mechanisms that underlie it. The provision of government subsidies can boost permanent income and reduce incentives for full labor market participation. Labor market withdrawal can confer benefits on those who fully retire, such as less stress, improved diet, and better sleep. However, it can also generate unintended adverse effects. For example, reduced labor market activities could decrease social engagement and mental acuity. As a result, the net effect is theoretically ambiguous. Although we show that the NRPS improved various health behaviors, we also demonstrate that the program led to a substantial reduction in social engagement, volunteering, and activities involving mental acuity. Overall, in terms of cognitive performance, the adverse effects on the mediating input factors likely outweigh the positive effects.

It is important to place our findings in the context of prior empirical estimates. Using data from high-income countries (in the European Union and the United States), Rohwedder and Willis (2010) and Mazzonna and Peracchi (2012) investigate the effect of early retirement on memory performance. Both studies document considerable harmful effects on cognitive performance associated with early retirement, a phenomenon Rohwedder and Willis (2010) refer to as *mental retirement*.[4] The estimated effect size in Rohwedder and Willis (2010) is more than one standard deviation for people in their sample. Compared to the estimates in that study, our estimates of the negative effects on cognitive performance are much smaller.[5]

[4] Using data from the U.S., England, Canada, and 11 European countries, Rohwedder and Willis (2010), Bonsang et al. (2012), and Adam et al. (2007) examine how retirement rates influence cognitive functioning and find a significant negative effect of retirement on cognitive functioning. Conversely, Coe et al. (2012) find no conclusive evidence using data from the U.S. Health and Retirement Survey (HRS). Other recent studies also examine the effect of the NRPS on other individual and household-related outcomes; Nikolov and Adelman (2009ab), for instance, examine how the NRPS affected intergenerational transfers and health behaviors.

[5] Mazzonna and Peracchi (2012) find a negative effect of retirement on orientation, immediate recall, and numeracy skills. The effect size is between approximately 0.2 to 0.3 standard deviations of the raw baseline cognitive performance measures (considerably larger than the estimates from our analysis). The relationship between retirement and cognitive decline has also been studied using data from high-income countries (Adam et al. 2007, Rohwedder and Willis 2010, Bonsang et al. 2012, Coe et al. 2012, Mazzonna and Peracchi 2012, Bingley and Martinello 2013, de

Our study contributes a new angle to the existing literature on participation in retirement programs in low- and middle-income countries (LMICs). First, we are among the first studies to examine how access to a retirement plan affects cognitive performance in the context of a developing country, and our study relies on a rich new dataset supplemented by analyses of administrative records.[6,7] Illuminating how retirement programs can generate adverse downstream effects on old-age cognition can provide insights for enriching existing models on human capital depreciation. Furthermore, from a policy standpoint, uncovering the potential mechanisms that lead to old-age cognitive decline can inform debates on creating policies to mitigate some of the adverse impacts without engendering the numerous benefits that retirement programs can confer to beneficiaries. Studying the depreciation of human capital is especially relevant in China because of the country's population size and the growing share of its elderly population. Second, we show how program participation affects a broader set of cognitive domains than has been previously considered. This study uses data on various proxies of cognition, such as episodic memory, sensitive to the aging process. As we age, episodic memory

---

Grip et al. 2015). Additionally, we discover that the provision of pension benefits has a greater influence on delayed recall than other cognition measures. The "delayed recall" test is one of the most sensitive tests to distinguish the effects of normal aging from Alzheimer's disease (Laakso et al. 2000). We rely on data from a rural sample in a developing country, whereas Rohwedder and Willis (2010) and Mazzonna and Peracchi (2012) use data from high-income countries; this discrepancy  is a significant factor that could explain some of the discrepancies among these studies.

[6] Recent studies examine the effect of retirement policies on health behaviors in the context of high-income countries (Eibich 2015) or developing economies (Nikolov and Adelman 2019b). Nikolov and Adelman (2019b, 2021) show that older adults with access to the NRPS pension program experienced significant improvements in several health measures.

[7] Cheng et al. (2018) assess the health consequences of the NRPS using data from the Chinese Longitudinal Healthy Longevity Survey, which was carried out a year after the introduction of the NRPS. Moreover, Cheng et al. (2018) rely on only one year of survey data after the NRPS introduction; the study does not directly observe NRPS participation (Cheng et al. 2018, pp.57). There is an overlap between the health inputs reported in Cheng et al. (2018) and the inputs in the cognitive depreciation process, which is the focus of this study. We revisit the issue in Section V as we delve deeper into the potential mechanisms underlying cognitive deterioration among the elderly induced by their access to the NRPS.

is the first domain to deteriorate (Souchay et al. 2000; Prull et al. 2000).[8] Finally, we provide insights on possible mechanisms underlying the observed impact of pension benefits on cognitive functioning.

The remainder of the study proceeds as follows. Section II outlines the implementation of the rural pension scheme and summarizes the data. Section III presents the identification strategy. Section IV presents the results. Section V reports additional robustness checks and bolsters the validity of the empirical approach. Section VI concludes.

## II. Background on the Pension Program and Data

### A. The New Rural Pension Scheme

*NRPS History and Expansion.* Based on a 2008 pilot project initiated in the city of Baoji in the northwestern province of Shaanxi, the Hu-Wen administration proposed an ambitious transformation of its pension system, launching the NRPS in 2009. Rural residents over 16 who did not already engage in an urban pension plan were eligible to join the NRPS on a voluntary basis. The rollout occurred in administrative areas based on the rural Hukou registration system.

The NRPS aimed to achieve full geographical coverage in rural areas by 2020 (Dorfman et al. 2013). Program coverage expanded between 2009 and 2013 (as shown in Figure 1), at 23 percent of districts (or 29 million beneficiaries) by the end of 2010 and to over 60 percent of districts (or 134 million beneficiaries) by early 2012.[9,10] By the end of 2011, over 50 percent of rural residents had contributed to the NRPS. Total participation in the program grew from 87 million to 326 million people from 2009 to 2011.[11,12]

[8] We use data from the CHARLS, a survey harmonized with the U.S. Health and Retirement Study (HRS) and other sister health surveys in high- and middle-income countries. The survey's harmonization of cognition measures across surveys enables additional international comparisons of human capital depreciation patterns across countries.

[9] Appendix Figure A1 shows the fraction of the senior population (within a community) receiving NRPS benefits over the four years.

[10] The central government of China released criteria for program implementation at the county level, and it aimed to distribute the first wave of participants equally among selected counties. The NRPS was launched earlier in counties in the middle and western areas throughout the course of the following two years by the central government. In Section

[Figure 1 about here]

*Program Eligibility, Participation, and Annuity Benefits*. Individuals who opted into the NRPS contributed towards benefits that they would be entitled to upon reaching 60. The newly launched program extended grandfathering conditions for residents who had already reached age 60 when the program launched. These individuals were eligible to receive a standard monthly benefit of 55 RMB[13], provided they had children who paid contributions towards the program. Participants aged 45 to 60, who had fewer years left to make contributions, had higher monthly contributions to compensate for the delayed participation. Individuals aged 60 or older can start collecting their basic pension every month without making any contributions if their children living in the same village participate in the NRPS. Individuals aged 45 to 60 can receive a basic pension after age 60 if they contribute each year until they reach 60. Those under age 45 are eligible to receive a basic pension after age 60 if they have paid into the system for at least 15 years.

Only people who contribute to the pension program are entitled to benefits, which comprise two components: (1) a basic pension of at least 55 RMB per month and (2) individual

---

III.A, we discuss the participant characteristics and show how similar these are between NRPS-implementing and -non-implementing areas.

[11] We analyze whether the age makeup of communities that adopted the NRPS earlier differed from the demographics of communities that did so later. There is no evidence that the average age between these two groups differs. The NRPS-participating areas and non-participating areas, however, could differ in other socioeconomic dimensions, an issue which we address in Section III.

[12] Three major factors account for the expansion of the NRPS between 2009 and 2011. First, the high rate of economic growth in these years played a considerable role. Between 2009 and 2011, the economy grew at an average annual rate of 9.3 percent, which provided robust fiscal capacity for the massive rollout of the program. Second, because of increasing income inequality and demographic pressures, demand for the program was substantial. Third, pension reform and program expansion into rural areas were fundamental priorities for the Hu-Wen administration.

[13] Nine percent of our sample households earn less than 55 RMB a month in 2011. In addition, 55 RMB is a quarter of a household's monthly income for 26 percent of the households in our sample.

account funds based on individual contributions and government subsidies. Individuals can typically opt for one of five annual contribution levels: 100, 200, 300, 400, or 500 RMB. These levels of voluntary contributions range from two to eight percent of China's rural yearly per capita net income in 2009.[14] The entire amount that can be paid out after 60 depends on earlier contributions and matching payments from the local government. Local governments annually provide at least 30 RMB in matching subsidies in addition to individual contributions.[15] Based on the chosen contribution level, the government subsidizes individual contributions (e.g., a government subsidy of 30 RMB/year for a contribution level of 100 or 200 RMB/year; a government subsidy of 40 RMB/year for a contribution level of 300 RMB/year; a government subsidy of 45 RMB/year for a contribution level of 400 RMB/year). According to the People's Bank of China, the interest rate on the account is the one-year base rate, approximately 2.5 percent in 2011. Changes in the base rate result in adjustments to the interest rate, which is compounded annually. Based on data collected from early program implementation, nearly 50 percent of participants opted for a minimum annual contribution of 100 RMB (Dorfman et al. 2013).[16]

---

[14] A participant may stop contributing for a few years and make up for the missed contributions later; they only lose the subsidies that they could receive if they contributed for the time they did not make program contributions. Partial withdrawals from retirement accounts are not permitted. Participants can withdraw all of their savings only under the following conditions: migration, change from a rural hukou to an urban hukou, or enrollment in an urban pension plan.

[15] This amount is independent of the individual contribution amount and may be higher depending on the local government's budget. This match amount is less than one-to-one, given the minimum contribution is 100 RMB and the basic match is 30 RMB. Lei, Zhang, and Zhao (2013) show that most participating municipalities, as of 2012, contributed 30 RMB per person per year.

[16] Lei, Zhang, and Zhao (2013) conduct various simulations on the present value, factoring in the opportunity cost of accumulated pension accounts and the present value of the accumulated benefit using the current (at the time of the study) rate of return for the NRPS program. They show that the most optimal investment strategy – if individuals maximize the net present value of the NRPS contributions – is choosing the lowest premium level and contributing as late as possible. They show that when the subsidy is 30 RMB/year, the optimal option is to contribute less than 21 years and choose the lowest premium level (100 RMB). This strategy has a positive net benefit (under additional assumptions about the annual interest rate, the timing when the pension benefits are claimed, and the annuity factor). The return on contributions above the lowest contribution levels is limited to the return on pension assets (i.e., the one-year bank deposit rate). Therefore, participants have a strong incentive to opt for the lowest contribution level

Participants receive benefits based on the "139 Rule," which divides the accumulated balance in the individual account over 139 months, the average remaining life expectancy in China at age 60.[17] Thus, the monthly payment comprises the basic pension plus the individual account balance divided by 139. Pension payouts do not depend on an earnings test, so participants can continue to work after receiving their pension income if they wish to do so.

Except for the local government subsidy, the individual account is completely inheritable upon the recipient's death. Pensioners' heirs receive a lump sum payout for the remaining balance in the individual account, excluding any government subsidies if the pensioner passes away earlier than the projected 139 months following them reaching age 60. If pensioners live more than 139 months after age 60, they still receive a monthly pension annuity until death.

In the rest of this section, we describe how we construct our analysis sample, define the variables we use, and present summary statistics.

## B. Sample and Cognition Measures

Our primary analysis draws on the CHARLS, which collects data on NRPS participation and cognitive performance among adults aged 40 and over. The second part of our analysis draws on the China Health and Nutrition Survey (CHNS), which we use because it collects information from the period before the launch of the NRPS.

*The China Health and Retirement Longitudinal Studies (CHARLS).* The CHARLS is a nationally representative survey that collects information on households with at least one person who is 45 years or older.[18] The sampling frame comprises all Chinese provinces and counties, except for

(100 RMB). The matching government subsidy of 30 RMB for an annual contribution of 100 RMB is likely too modest to incentivize workers to contribute beyond the 15-year vesting period for the basic benefit.

[17] The individual account has a rate of return equal to the People's Bank of China's one-year deposit rate. The "139 Rule" was adopted from the previously established Urban Pension Scheme.

[18] For context, life expectancy at birth in China was 74.5 years for men and 76.7 years for females in 2018 (UN 2019). The typical retirement age in rural areas is 61 years (SOA 2016). The average household income (excluding pension

Tibet. The CHARLS collects demographic data on family structure, cognition, health, pensions, retirement, work, household wealth, income, and consumption. Data collection started a year after the launch of the NRPS.

We construct our analysis sample from individual-level panel data. The sample comprises 15,990 individuals across two waves of the CHARLS, who come from 429 communities across 28 provinces. The raw sample totals 17,708 individuals (constituting 10,287 households) from 450 communities across 28 provinces. The 2011 baseline wave interviewed 10,257 households with 18,245 respondents aged 45 and over.[19] The 2013 follow-up wave covered 10,979 households (or 19,666 respondents). The CHARLS collects individual participation in various government programs, including the NRPS. We dropped observations with urban *hukou* status for our primary analysis sample. Individuals attached to an urban *hukou* are ineligible to participate in the NRPS and instead participate in urban pension schemes. Our analysis also does not include individuals over the age of 60 with no children at the time of program coverage because they were ineligible to participate in the NRPS.

Table 1 reports the summary statistics for our sample.[20] Among participants and non-participants, 70 percent and 69 percent were employed in the baseline wave, respectively. Seventy-two percent of participants and non-participants work in agriculture. The rural sample reported low educational attainment levels—approximately 46 percent to 48 percent report

---

income) was 26,022 RMB; the median household income was 11,680 RMB. On average, NRPS beneficiaries received 233 RMB/year (i.e., approximately 1 percent of the average income or 2 percent of the median income).

[19] Appendix Figure B1 shows the geographical coverage map for the CHARLS survey. Some 19,081 households were initially sampled, but only 12,740 of them had age-eligible members, of which 10,257 responded.

[20] A potential issue related to the sample used in our analysis is *sample selection*, which refers to the possibility that the analysis sample may differ from the underlying survey sample. In Table 1, we report the characteristics of the analysis sample (columns 3 and 4), and in column 2, we report the characteristics of the full sample. If sample selection were an issue, we would have observed a large difference between the underlying source sample and the analysis sample (i.e., individuals who live in rural areas). Furthermore, we benchmark the data from Table 1 to other published studies (e.g., Caro and Parada-Contzen, 2021), and we do not detect significant differences between the characteristics of the rural sample that we use, and the characteristics observed in these studies.

having completed at least a secondary school degree. Twenty-seven percent of participants and 26 percent of non-participants report at least "poor/fair" health.

[Table 1 about here]

*Cognition Measures.* The CHARLS also includes measures of cognitive performance (Ofstedal 2005).[21] The first cognition measure is episodic memory, captured via verbal learning and several recall tasks.[22] The second cognition measure is mental intactness; the task involves recognition of date (month, day, year, season, day of the week), self-rated memory (excellent, very good, good, fair, and poor), and serial subtraction of 7s from 100 (up to five times). The survey asks each respondent to redraw a picture of overlapping pentagons. We use the sum of the two scores for the immediate and delayed recall measures (both tests are part of episodic memory)— the total score ranges from 0 to 20. Low scores on this total word sum are indicative of low memory capacity. We analyze all cognition measures tested in the survey. However, we pay special attention to the episodic memory domain, as several studies point to this domain being critical for cognitive aging (Souchay et al. 2000; Prull et al. 2000). In addition to the cognitive tests, respondents rate their memory based on a 5-point scale. The 5-point scale used for the measurement of episodic memory is as follows: (1) Excellent (2) Very Good (3) Good (4) Fair (5) Poor.

[21] In general, and based on Cattell's psychometrically-based theory, intelligence comprises two components: fluid and crystallized intelligence (Brown 2016). Therefore, we can situate the CHARLS proxies of episodic memory and mental intactness in the framework of these two general domains (fluid intelligence and crystallized intelligence). Drawing conclusions, recalling memories, and abstract reasoning are all parts of fluid intelligence. Fluid intelligence is used, for instance, when solving new challenges and thinking on the fly. On the other hand, accumulated knowledge makes up crystallized intelligence. Cryptized knowledge is acquired via education and practical experience. Our first measure, episodic memory, includes the capacity to reason and recall details from memories, thus it captures some components of fluid intelligence (McArdle et al. 2011). In contrast, the mental intactness test, which is related to determining and accessing the knowledge stockpile known as human capital, captures both fluid and crystallized intelligence (McArdle et al. 2011).

[22] HARLS uses the HRS version of the CERAD immediate and delayed word recall tests to assess episodic memory (Ofstedal et al. 2005). *Episodic memory* is a necessary component of reasoning in many dimensions. The two tasks that capture verbal learning and recall are immediate and delayed recall. After approximately four minutes after other questions, the respondent is asked again to list the nouns, without reading the words a second time. Word recall tests were collected to assess the severity of short-term and long-term cognitive impairment. For the immediate recall test, surveyors randomly assign respondents to a list containing ten common words. The respondent is given two minutes to recall as many words as he/she can remember. The immediate recall score ranges from zero to ten and provides the number of words remembered correctly. Following the recall, the respondent answered unrelated questions for several minutes until prompted to recall the original word list. This procedure captures the delayed recall score, which ranges from zero to ten.

We also report impacts on the following outcomes: perceived memory status (subjective status), knowing the current month (orientation), serial-7 score (working memory), immediate recall score (memory capacity), and delayed recall score (memory duration). We aggregate these multiple measures into one composite index.[23,24] We show, in Table 1, data on the cognitive measures of individuals in our sample. NRPS participants scored slightly better on the word recall tests. The average score on the immediate word recall task was 3.93 out of 10 (non-participants average 3.77 out of 10). Similarly, delayed recall scores were higher for participants than for non-participants at 2.91 and 2.89. Approximately 84 percent of participants and non-participants correctly named the current month. Based on the aggregated index measure, the cognitive index exhibits a higher average for participants than for non-participants, 0.06 and 0.00, respectively.

*China Health and Nutrition Survey.* The CHARLS does not collect cognition data before the start of the NRPS. Therefore, we draw on a second source, the China Health and Nutrition Survey (CHNS), to examine some of the underlying assumptions of our main empirical strategy.

[23] We use a method based on Anderson (2008) to transform the set of proxy variables for cognition into an aggregated index. We do so by first standardizing each cognition proxy. We then compute the covariance matrix for all cognition measures. Third, we compute the eigenvectors; combined, they contain the same information as the original variables. The first component, based on the largest eigenvalue, contains the most information by design, whereas the last component contains the least. We reduce the original cognition proxies into one index by retaining the component with the largest variance (eigenvalue). The overall index, Cognitive Memory Index, has a mean of zero and a standard deviation of 1.42. The index provides a normalization of cognitive memory status, where negative (or low) values are associated with poor memory functioning. This index is an overall cognition proxy in the analyses that we present in the next section.

[24] To address potential multiple hypotheses testing issue (for the cognition proxies and other outcomes), we employ an indexing approach, as advocated in recent methodological studies (Viviano et al. 2021, Clarke 2021). As a robustness check (not reported here), we also estimate adjusted p-values based on the Westfall-Young procedure (a less conservative version of the Bonferroni adjustment). The results and the pattern remain robust to these additional estimation procedures.

The CHNS covers 1989 to 2011, a period overlapping with the start of NRPS. The CHNS sampling areas overlap with those sampled by the CHARLS.[25] The survey covers approximately 19,000 individuals in 15 provinces spanning 216 primary sampling units (PSUs).[26,27]

The CHNS also collected information on proxy measures of memory and cognition, similar to those collected by the CHARLS (UNC-Chapel Hill 2010). The CHNS adopted similar cognitive screening items because it adopted measures from the United States-based HRS survey. The cognition tests tested immediate and delayed recall of a two-word list, counting backward from 20, serial-7 subtraction, and memory orientation. Scores for immediate and delayed recall ranged from zero to ten. Counting backward and serial 7s were used to assess attention and calculation, with scores ranging from zero to seven. Orientation was assessed by asking the participant for the current date (one point each for a correct response on the year, month, and date) and the name of a tool used to cut paper (one point). Higher scores across the board indicate improved cognitive performance.

## III. Empirical Strategy: NRPS Implementation and Program Impacts

### A. Triple Difference Estimation

Our primary goal is to investigate the effects of the NRPS rollout on cognition among program beneficiaries. Our identification strategy relies on variation across municipalities based on their NRPS implementation.[28]

---

[25] Figure B1 depicts the geographical coverage of the CHNS. Appendix Table B1 compares the summary statistics for the CHARLS and CHNS for adults ages 45 and over and living in rural areas. CHARLS only collects information on adults ages 45 and over.

[26] The survey covered the following provinces: Beijing, Chongqing, Guangxi, Heilongjiang, Henan, Hubei, Hunan, Jiangsu, Liaoning, Shaanxi, Shandong, Shanghai, Yunnan, and Zhejiang. The CHNS collected data on fourteen provinces in contrast to the twenty-eight provinces in the CHARLS.

[27] When we examine sample characteristics for identical locations, periods, and age groups, we note that the characteristics of the survey samples are quite similar.

[28] Figure B2 details the distribution of individual NRPS participation at the community level.

We estimate the effects of the NPRS using a DDD estimator. The estimator uses variation based on community participation (some communities implemented the program, and some did not), the timing of the NRPS implementation (some areas adopted the program earlier than others did), as well as the pension annuity's age eligibility condition.[29] Although the identifying variation comes from the treated areas (i.e., communities) between 2011 and 2013, we analyze the longitudinal data at the individual level. Based on information from the CHARLS, we construct a variable, $OfferNRPS_{ct}$, which indicates the participation status (whether a community *c* adopted the NRPS program at time *t*). The CHARLS records the place of residence, whether the local municipality implemented the NRPS, and whether respondents enrolled in the NRPS.[30] This data aspect allows us to define the variable, $OfferNRPS_{ct}$, based on responses from individual-level data. In its early waves. In its early waves, the CHARLS does not directly measure community implementation of the NRPS. Therefore, we adopt the following approach: if no individual indicates to have NRPS at time *t* in community *c*, then $\text{OfferNRPS}_{ct}$ is coded as 0; if at least one person reports participating in the NRPS, $OfferNRPS_{ct}$ is set to 1.[31] We examine the impact of NRPS provision on cognition using the following specification:

$$(1)\qquad Y_{ict} = \beta_0 + \beta_1(OfferNRPS_{ct} \times Above60_{ict}) + \beta_2 Above60_{ict} + \beta_3 \boldsymbol{X}_{ict} + \phi_c + \mu_t + \phi_c \times Above60_{ict} + \mu_t \times Above60_{ict} + \phi_c \times \mu_t + \varepsilon_{ict},$$

[29] In Online Appendix Table A1, we show that the characteristics of the participants in NRPS implementing and non-implementing areas are similar.

[30] It is not common for individuals to stop participating in the NRPS once they are enrolled in the program; however, 705 individuals indicated they were enrolled in 2011 but not in 2013. At the community level, eight communities (out of 350 communities used in our sample) implemented NRPS in 2011 but not in 2013. Only 7.5 percent of the 705 individuals who stopped participating in the NRPS resided remained n the eight communities that indicated terminating their NRPS implementation. Therefore, these individuals are not a threat to the internal validity of the community-level instrumental variable approach.

[31] We address concerns regarding measurement error and any potential associated bias in the estimated coefficients, and we report additional robustness checks in Section V.

where $Y_{ict}$ is the cognition outcome and $Above60_{ict}$ is equal to 1 if the respondent is aged 60 and over (and 0 otherwise). $\boldsymbol{X}_{ict}$ is a vector of individual-level controls; $\phi_c$ and $\mu_t$ are community and time-related fixed effects. We include community fixed effects and community-time fixed effects, $\phi_c \times \mu_t$ to control for community differences during NRPS implementation.[32,33,34] We estimate specification (1) with and without individual fixed effects.

The coefficient of interest in (1) is $\beta_1$. It captures the intent-to-treat (ITT) estimate of the average effect of the NRPS program on the average outcomes of eligible individuals aged 60 and over who live in a treated community, regardless of whether the individual decides to participate in the program. Because it takes into account potential region-specific effects and has been used to evaluate similar policy rollouts, the DDD design is the best option to study the program impacts (see Katz 1996). We recover the NRPS treatment effects using the DDD design instead of using the DD approach. The DDD approach can address potential confounding trends in a way that the DD design cannot: (1) changes in the cognitive performance of individuals over 60 in the communities that implemented the NRPS, and (2) changes in the cognitive performance of all people who reside in NRPS-implementing communities.

If the variation in program implementation across communities is unrelated to other community-related shocks, specification (1) will produce an unbiased estimate for $\beta_1$, our main coefficient of interest. The DDD design also rests on an assumption that the cognitive measure gap between people in the two age groups (i.e., 60 and over versus less than 60 years) would have evolved similarly in the NRPS-implementing communities as it would have evolved in the comparable control communities (Cunningham 2021, Olden and Moen 2020). To check whether the triple difference approach is an appropriate empirical strategy, we test the common trends assumption using pre-policy survey data, following the approach adopted in Autor (2003). Using data before the 2009 implementation of the NRPS, we analyze the trends of various cognition measures in treated and untreated areas. We can only conduct pre-trends analysis with data from the CHNS using waves that collected cognition measures before 2009 because the CHARLS collected survey data after the NRPS's inception.

[32] In several robustness exercises, we also include a richer set of age-related controls (e.g., age-squared), and results remain robust in these specifications. We cannot include age directly as it is directly collinear with the binary indicator for being over age 60. We cluster the standard errors by community and age groups. We also estimate the fully saturated triple difference specification.
[33] In Online Appendix B, we report additional robustness checks where we cluster the standard errors by community and age. Our results are robust to community and age-specific clusters.
[34] In a related study, which explores the impacts of NRPS on inter-household transfers, Nikolov and Adelman (2019b) rely on a Tobit model analysis, in contrast to the DDD approach employed in this paper.

The primary challenge in using the CHNS is that community identifiers and geographic-level variables do not match one-to-one with those in the CHARLS. Due to these data limitations in the CHNS, we cannot identify the timing of the NRPS implementation at the county level. Only for this empirical exercise do we redefine "treated" and "control" units at the province level (as opposed to the community *shequ* level). The CHNS collects data on cognitive performance only from the 2000 wave onwards. Therefore, we test the common trends assumption at the province level using the CHNS from 2000 to 2009. We use a binary variable for treatment status in each province, defining a province as "treated" (=1) if more than 67 percent of all communities within the province implemented the NRPS (according to the baseline wave of the CHARLS) and "control" (=0) otherwise. We also conduct additional sensitivity analyses using alternative threshold choices. Specifically, we re-estimate our specifications with a lower (50 percent coverage rate) and a higher (70 percent coverage rate) threshold choice.[35] Our analysis

[35] CHNS does not sample from the same communities/villages as the CHARLS. Therefore, we rely on our definition of treated and control provinces based on the CHARLS to test data in the CHNS. We report the results of these sensitivity analyses in Appendix Table A2.

for this test uses data on cognition outcomes from the CHNS that mirror the cognition measures collected by the CHARLS. Using the CHNS data on cognition measures from the 2000, 2004, and 2006 waves, we estimate the following specification:

(2) $$Y_{ipt} = \beta_0 + \beta_{-3} D_{ipt-3} + \beta_{-1} D_{ipt-1} + \phi_p + \mu_t + \phi_p \times \mu_t + \varepsilon_{ipt} \, ,$$

where $Y_{ipt}$ is the cognition proxy, and $\phi_p$, $\mu_t$, and $\phi_p \times \mu_t$ are province, time, and province-time fixed effects, respectively. Because of the triple-difference estimation, we include the triple interactions $D_{ipt} = (OfferNRPS_{pt} \times Above60_{ipt})$ for the first and last pre-treatment periods.[36] $D_{ipt}$ is defined in the same way as in our main triple-difference specification; subscript $p$ denotes the province. The results reported in Online Appendix Table A3 provide clear evidence that $\beta_{-3}$ and $\beta_{-1}$ are insignificant. Based on this test, we fail to reject the hypothesis regarding similar trends for outcomes between NRPS and non-NRPS areas.[37, 38]

Furthermore, because we measure outcomes at the individual level, we conduct the same exercise but adjust the definition of treatment status for treated areas by each community's population size. We define treatment status based on the percentage of individuals (as a fraction of all individuals in that community) who report participating in the NRPS: if 10 percent or more report participating in the NRPS, we code the community as participating in the NRPS. We then aggregated the information from participating communities to the province level, as previously described. We perform the same population-based exercise as a robustness check by increasing the threshold of defining a treated municipality to 20 percent, 30 percent, 50 percent, and 60 percent. Online Appendix Table A2 reports the results. Again, the results show no evidence that the estimates for $\beta_{-3}$ and $\beta_{-1}$ are statistically significant. Therefore, this exercise provides no empirical support refuting the validity of the common trends assumption.

[36] In this specification, the second pre-treatment period is omitted.

[37] This test is based on analysis at the province level, as described above. The CHNS only covers data from nine provinces. Therefore, the number of observations only in this analysis sample is considerably smaller than in the primary sample (at the community level). The low number of observations may lead to an underpowered inference for this test. However, Table A3 also shows that the estimated coefficients are statistically insignificant and unstable (i.e., the effect size differs substantially from the primary estimations) across different threshold specifications. These empirical estimates further undermine any evidence of robust differences, at the province level, in pre-trends.

[38] As already noted, we conduct the pre-trends analysis using the CHNS data at the province level because of the data limitations. The regressor in equation (2) is the so-called generated regressor. Murphy and Topel (1985) note that generated regressors have a sampling variance of their own, and such estimations tend to produce standard errors that are too small. However, our empirical test based on equation (2) produces non-significant results. When we bootstrap the estimation procedure, the standard errors become even larger.

Finally, we investigate a potential threat—selective migration— to the internal validity of our empirical approach. If the factors affecting migration correlate with the timing of the implementation of the NRPS, then our estimates could be biased. Related to this issue, Dou and Liu (2017) investigate the volume of internal migration in rural China. They find that, between 2011 and 2013, only 6.1% of older people moved locally; only 0.5 percent made a long-distance move.[39] Second, when we examine the timing of the NRPS roll-out in relation to migration decisions, we discover little indication that there is a connection between the two factors.[40]

### B. Instrumental Variable Estimation

To explore the possibility of endogenous individual participation, we augment the DDD analysis by using $OfferNRPS_{ct}$ to instrument for individual participation in the NRPS, following an instrumented difference-in-differences design as in Hudson, Hull, and Liebersohn (2017).[41]

[39] We find a similar pattern in our sample: about 97.4 percent of rural seniors live in the county they registered in at birth.

[40] We conduct an exercise in which we regress a binary indicator if someone migrated between 2011 and 2013 on a set of covariates, including a binary indicator of whether an area implemented the NRPS. We report the results in Online Appendix Table A4: we find no evidence that cross-county migration decisions correlate with whether an area adopts the NRPS.

[41] Hudson, Hull, and Liebersohn (2017) note the identifying assumptions for the DDD-IV estimation: the typical exclusion restriction for an IV, parallel trends (growth paths of both treatment and outcomes need to be independent of the actual instrument assignment), and monotonicity (the effect of the instrument, the NRPS availability, on the NRPS take-up, needs to be monotone).

Here, we define $OfferNRPS_{ct}$ as an indicator variable that is set to 0 if no individuals participate in the NRPS and set to one if there is at least one participant excluding the individual $i$'s response.[42] We re-estimate the following specification:

$$(3) \qquad Y_{ict}=\beta_0+\beta_1(\widehat{NRPS}_{ict}\times Above60_{ict})+\beta_2 Above60_{ict}+\beta_3\boldsymbol{X}_{ict}+\phi_c+\mu_t+\phi_c\times Above60_{ict}+\mu_t\times Above60_{ict}+\phi_c\times\mu_t+\varepsilon_{ict}.$$

$\widehat{NRPS}_{ict}$ represents individual enrollment in NRPS, which we instrument with $OfferNRPS_{ct}$. $\boldsymbol{X}_{ict}$ is a vector of individual-level controls. $\phi_c$, $\mu_t$, and $\phi_c\times\mu_t$ are community, time, and community-time fixed effects, respectively. The community fixed effects capture time-invariant community characteristics.

# IV. Program Impacts

## A. Program Access and Cognitive Performance

We begin our analysis by examining the impacts on cognitive outcomes from estimating specifications (1) and (3). Table 2 reports the results, with and without individual fixed effects. Columns 1 through 8 report the estimated effects on the various cognition proxies for immediate recall, delayed recall, total recall, and memory index. These results report estimates based on equation (1); therefore, they are the intent-to-treat estimates of the effect of program availability in a community on cognition measures.[43,44] Panel B reports the treatment-on-treated estimates from the 2SLS approach, with and without individual fixed effects.

[42] The CHARLS does not collect information on the NRPS implementation at the community level.

Both tables indicate a striking pattern of adverse effects of NRPS availability on all cognition measures. On the immediate recall test, individuals in NRPS program areas aged 60 and above score, on average, worse by eight percent of a standard deviation. For delayed recall tests (with individual fixed effects), individuals in program areas score worse by approximately 14 percent of a standard deviation. Program availability also has a considerable negative impact on the cognitive index. As described in Section II, this index combines the cognition measures of mental intactness. On average, the provision of NRPS benefits leads to a 0.12-point reduction in the composite score (equivalent to about 9 percent of a standard deviation) in the intent-to-treat specifications. The treatment-on-treated estimates for specification (3) reported in Panel B are approximately double the size of the ITT estimates. As with the results reported in Panel A, the 2SLS estimates are statistically significant and negative for all cognition measures. Comparing the estimates across all columns, we find the most substantial adverse effect on the delayed recall cognition measure, at approximately double the effect size for the two other cognition measures. Neurological studies document that this specific proxy measure of cognition is a helpful predictor of dementia in adulthood (Welsh et al. 1991, Laakso et al. 2000).

[Table 2 about here]

---

[43] We also perform a simple DD analysis for two subgroups: people aged 60 and above who reside in rural areas and people aged below 60 who live in rural areas. We find that the NRPS does not influence the cognitive performance of age-ineligible people living in rural areas.

[44] The number of observations used in the analysis (reported in Table 2) differs from the number of observations in the baseline sample due to two main reasons: small sample attrition (approximately 8 percent between the first two waves) and the addition of a refreshment sample in the follow-up CHARLS waves. We examine whether there is a relationship between attrition status and the treatment variable (i.e., access to NRPS benefits at the community level). We find no evidence of differential attrition (see Online Appendix Table A5).

Thus far, we have only reported results based on access to the NRPS program. It is pertinent to note that the NRPS is not an employment-based program; the receipt of benefits is independent of employment status. The program could induce early labor market withdrawal, an issue we return to in the next section, where we shed light on potential mechanisms. To explore whether changes in cognitive performance operate through a change in retirement status, we repeat specifications (1) and (3) but instrument retirement status as the treatment. Using the formal definition of retirement and other employment-related variables available in the CHARLS, we reconstruct the definition of retirement.[45] Table 3 reports results from estimating specifications (1) and (3) using this reconstructed variable. However, we recognize that the data for this particular variable is quite limited and probably prone to data quality problems.

[Table 3 about here]

The results, detailed in Table 3, show the detrimental effect of retirement on cognitive performance. Although not all results pass the conventional levels of statistical significance, all effect size estimates are negative—columns 1 through 6 report negative coefficient estimates for the cognitive performance outcome.

Next, we analyze whether beneficiaries of the program experience a greater degree of cognitive decline depending on how long they have been exposed to NRPS benefits. The duration of receiving benefits should result in a bigger cognitive decline if the NRPS is the main cause of the cognitive decline. To examine this possibility, we categorize all program beneficiaries aged 60 or older into three categories: less than one year of exposure to NRPS benefits, between one and three years of exposure to NRPS benefits, and more than three years of exposure to NRPS benefits. We estimate the effect of length of exposure to NRPS benefits using the primary DDD estimator interacted with dummies for the first and third of these three categories. Table 4 reports the results. The reference group comprises beneficiaries with less than one year of exposure to NRPS benefits. Based on results reported in the fifth row, people who receive NRPS benefits for longer than three years show more pronounced cognitive decline than people who are exposed for a shorter amount of time. As a result, we offer empirical evidence that lends credence to the idea that NRPS claimants who get payments for a longer amount of time experience a more severe cognitive deterioration.[46]

[45] This re-constructed binary definition of retirement is based on available data for the following CHARLS variables: the person completed retirement procedures in any survey wave, the reported number of days (or months, hours) worked is zero in three consecutive waves, the reported usual number of days (or months) per year is zero for three consecutive waves, the reported monetary retirement benefit is positive, the number of workdays missed for health reasons has been more than 300 per year for three consecutive survey waves, the reported year of retirement is before 2009, and the survey respondent indicated that the formal retirement is processed. The identifying assumption of this estimation is that the availability of pension program benefits affects cognitive performance solely through participation in the labor market.

[Table 4 about here]

We also examine heterogeneity in program effects by gender.[47] The results reported in Online Appendix Table A6 echo the negative effect of the NRPS program on cognitive measures. However, the coefficients are not statistically significant. As a result, we find no conclusive evidence on treatment effect heterogeneity by gender.

[46] In analyses not reported here, we repeat the same exercise with a continuously defined variable capturing the length of exposure to NRPS benefits. We find the same negative pattern.

[47] We do so by interacting the DDD estimator test in specifications (1) and (3) with a binary indicator for female beneficiaries. Online Appendix Table 6 reports the results of the analysis. Panel A reports the intent-to-treat DDDD estimates, whereas Panel B provides estimates from the 2SLS estimation approach.

## B. Channels: How Do NRPS Benefits Induce Cognitive Decline?

Gaining access to retirement program benefits likely has an influence on behavior, which in turn influences the depreciation of human capital. We examine various retirement-related outcomes to probe more deeply into potential mechanisms among NRPS beneficiaries.

Using the panel data available in the CHARLS on potential channels, we examine how NRPS participation affects four activity groups among program beneficiaries: (1) labor market activities, (2) mental stimulation, (3) social engagement, and (4) health behaviors, time use, and health care utilization. Tables 5 and 6 report the results. Suppose we do not observe any change in an outcome that could play a mediating role in influencing cognition. In this case, we take this to indicate that the causal pathway does not operate via that mediating factor (or group of factors).

Table 5 reveals that NRPS participation led to adverse impacts on labor market outcomes, mental stimulation, and social engagement. Among NRPS participants who remained active in the labor force, we see a pattern of reduced activity in the labor market (for self-employment activities); the results on wage employment are not statistically significant. Similarly, we find evidence of decreased mental stimulation among NRPS participants. However, the evidence of decreased mental stimulation is only suggestive as the point estimates are negative but statistically insignificant.

Finally, we find substantial evidence of decreased social engagement among NRPS beneficiaries (Columns 9-12 in Table 5). Participants in the program report substantially lower levels of social engagement, with significantly lower rates of volunteering and social interaction than non-beneficiaries. Related to this finding, recent empirical investigations focusing on the connection between social engagement and cognitive impairment in older Chinese adults (65 years and older) show that lower social engagement is associated with higher cognitive impairment (Zuho et al., 2020; Zhou et al., 2020).

The reported results above indicate that social isolation is strongly associated with cognitive decline among the elderly in rural China. We further explore this association by investigating whether the number of household members of the NRPS participants mediates the relationship: if social isolation is a critical factor for cognitive decline, NRPS participants who are part of larger families will exhibit attenuated cognitive decline. In Online Appendix Table A7, we break down the estimates into three family group sizes: up to 2 members, 3 to 5 members, and more than 5 members. The results show that being part of a larger family is associated with lower cognitive decline.

On the other hand, we find (in Table 6) that program participation improved some health behaviors. Program participants reported a reduced incidence of regular alcohol drinking compared to the previous year (Column 3 in Table 6). Program participants, on average, also smoked less. Although we are unable to further investigate this interpretation due to data constraints, these findings are consistent with the notion suggesting program participants are less stressed. Participation in NRPS enhanced sleep patterns as well. Therefore, in light of these empirical findings, it is unlikely that program members' health behaviors can explain the cognitive decline among elderly residents of rural China.

[Tables 5 and 6 about here]

In summary, we find that NRPS participation has many positive effects on diet, smoking, and other health behaviors. However, the program has also adverse effects on mental health and mental stimulation. Due to data limitations, there may potentially be more channels of influence that we are unable to investigate. Overall, given the significant cognitive decline reported among program participants and the analysis of mediating mechanisms, it seems likely that the adverse effects of NRPS on mental and social engagement significantly outweigh the program's protective effect on various health behaviors.

## V. Robustness Checks

We perform a series of robustness exercises. First, we conduct a falsification exercise to explore the validity of our empirical estimation. We re-estimate our primary specifications using a sample of individuals ineligible for the NRPS and see if we find insignificant program impacts. Second, we address the possibility of measurement error in how our analyses capture participation in the NRPS program. The main results survive these extension exercises.

### A. Falsification Exercises

We construct a falsification exercise based on an alternative sample of people who are not eligible for the NRPS or its benefits. In theory, when we rerun specifications (1) and (3), the coefficients of interest discussed in Section III for this alternative sample should not be significant.

As noted in Section III, the NRPS program is only available to individuals in rural administrative districts who are not already enrolled in one of the urban pension schemes. In the analysis in Section IV, we excluded urban pensioners and rural-residing elderly individuals without children because both groups are ineligible for the NRPS. For this falsification exercise, we reconstruct the analysis sample using these two groups: pensioners in an urban pension program or rural-residing individuals aged 60 and above who did not contribute to the NRPS before age 60 and have no children residing in rural administrative districts.

In this falsification exercise, we only perform the analysis based on a sample of individuals who: (1) live in rural areas but obtain benefits from an urban pension system; or (2) are elderly, without children and who happen to reside in rural areas. The main objective is to examine whether the primary estimation approach will detect cognitive performance effects among individuals in this placebo sample when such effects should not be present. If specifications (1) and (3) yield no spurious results, this falsification exercise should produce non-significant estimates for the coefficients associated with the NRPS effect on cognition outcomes.[48] As described in the primary empirical approach, we re-estimate specifications (1) using data from the placebo sample. We report the results in Appendix A, Table A8.

Table A8 reports non-significant program impacts on the immediate recall score, total recall score, and cognitive memory index. In other words, these results imply that the cognitive performance of urban pensioners who live in communities that offer the NRPS does differ from the cognitive performance of urban pensioners who live in communities not implementing the NRPS. These results further bolster the validity of the main results reported in Tables 2 and 4; they are unlikely to be based on a spurious specification choice. We also conduct an additional falsification test in which we re-estimate specifications (1) and (3) on a set of placebo outcomes. We select several placebo outcomes for which no plausible mechanism linking access to the pension program and treatment effects exists. We execute this additional falsification exercise as another attempt to examine the credibility of our main results. We use four placebo outcomes: a person's nationality being Han, the number of non-living daughters in the household, the mother's educational level, and the number of living sons. Appendix A Table A9 reports the results. Panel A reports the ITT results, and Panel B reports the TOT results. In both panels, the results provide no evidence of program effects on any of the placebo outcomes.

[48] This additional test assumes the absence of spillover effects between the group of individuals who are beneficiaries of the NRPS and the urban pensioners who live in the same communities that offer the NRPS. Still, they are not eligible (nor do they receive program benefits) for the NRPS. As we show in Section IV, NRPS participants have fewer social interactions. Therefore, if spillover effects occur via social interaction (a viable mechanism for social spillovers), then our analysis will pick up program impacts among non-beneficiaries who live in areas that offer NRPS, and we do not detect such changes.

## B. Alternative Measures of NRPS Participation

We also explore the possibility that our primary analysis relies on either mismeasured individual participation in the NRPS or incorrect reports among municipalities (*shequ*) of NRPS program implementation. Either of these possibilities will yield measurement error in our program impacts and could produce biased impact estimates. Therefore, we perform additional consistency checks based on alternative approaches intended to measure NRPS participation.

*Propensity Score Method Definition*. First, survey responses from the CHARLS may be incorrect, resulting in a possible mismeasurement of actual NRPS participation by individuals in our analysis. We address this possibility with an alternative measurement of individual NRPS participation status, using personal characteristics to reconstruct the likelihood of participating in the NRPS. We redefine the NRPS participation status based on a propensity score matching approach. We predict NRPS participation (at baseline) using a combination of individual characteristics, such as education, gender, parental education, and nationality. We use baseline data for these variables. Using these characteristics, we then predict the propensity of NRPS participation, $\widehat{NRPS}_{ic}$, based on the propensity score matching method. The predicted participation, based on this estimation technique, is $\mathrm{PrNRPS_{ic}}$. We construct an alternative measurement of the NRPS participation status variable by defining $PrNRPS_{ic}$ = 1 if $\widehat{NRPS}_{ic}$ is greater than one standard deviation above the mean of $\widehat{\mathrm{NRPS}}_{\mathrm{ic}}$.

We use this redefined measure of NRPS participation to re-estimate specifications (1) and (3). The NRPS program participation in this analysis uses the reconstructed variable ($PrNRPS_{ic}$), based on the estimation from the propensity score approach, as opposed to the estimation approach in Section IV (based on the self-reported variable in the CHARLS).[49] We report the results based on this alternative definition of program participation in Online Appendix Table B3. Using the alternative NRPS participation definition, the reported results bolster our main findings.

*Community NRPS Participation Definition*. We next consider the possibility of measurement error due to individuals misreporting their NRPS participation in the CHARLS. If true, this measurement error at the individual level could generate possible misclassification of areas reporting they implemented the NRPS, an issue affecting our analysis using specification (3).[50]

Thus, we verify the robustness of our approach with an alternative definition of the variable $OfferNRPS_{ct}$. This additional exercise is designed to remove possible contamination related to how we define what areas are *treated* (i.e., areas indicating NRPS implementation). We re-estimate specifications (1) but rely on a higher threshold that defines when the variable $OfferNRPS_{ct}$ (the variable that indicates community participation in the NRPS) switches from zero to one. Instead of relying on a threshold of at least one individual reporting NRPS participation to set $OfferNRPS_{ct}$= 1, we now use an alternative (and higher) threshold of at least four participants in community $c$ to set $OfferNRPS_{ct}$ being equal to 1. Furthermore, in yet another more stringent definition, we rely on a definition of at least seven individuals within the community participating in the NRPS regarding when the community indicator switches from zero (non-participating) to 1 (participating).

[49] Two key assumptions underlie this propensity score approach. First, the approach assumes that only observable (and time-invariant) characteristics determine selection for participation in the NRPS. Second, the method relies on the assumption that, in the absence of the NRPS, the age trend in cognitive functioning is the same between covariates, not used in our propensity score determination. In the matching step, we only include time-invariant variables, as argued in Imai, Kim, and Wang (2022).

[50] We conduct an additional extension exercise to address another potential source of measurement error in the variable that measures whether a community implements the NRPS. In the primary analysis, we define the NRPS program's implementation at the community level based on survey data at the individual level. In this empirical approach, if at least one individual in the community reports participating in the NRPS, we define the community as implementing the NRPS. However, communities with very few NRPS participants may be systematically similar in unobservable factors we cannot observe. This scenario could produce a measurement error affecting our measurement of the instrument. Therefore, in an additional extension exercise, we re-estimate by removing communities that report very few NRPS participants within their boundaries. We then proceed by re-estimating the main specifications reported in Section III. Online Appendix B Table B5 reports the results of this extension exercise. The pattern of the results remains consistent with the main results.

We report the results from these additional analyses in Online Appendix Table B4. The results demonstrate that our original estimates are robust to the alternative and more conservative definitions of the threshold, which determines when the variable $OfferNRPS_{ct}$ switches values.

*Using Online Administrative Data*. Using data from online sources, we probe the robustness of the results using administrative program announcements. For this empirical exercise, we comb data from Chinese newspapers (online or in paper format) based on public announcements regarding geographic participation in the NPRS. With administrative data, we can probe more deeply into how effectively we capture NRPS implementation; we can analyze NRPS implementation in the local municipality rather than relying on CHARLS data. Generally, public announcements are available at the city and community levels. Our focus is on the public announcements for NRPS implementation in Heilongjiang.[51] The Heilongjiang province is ideal for this empirical exercise for several reasons. First, online announcements regarding city-level implementation of the NRPS between 2011 and 2013 are readily available. As outlined in Section II, data is available from the two CHARLS waves for this period. Second, the province is one of the largest provinces in China. This factor can considerably facilitate the re-estimation exercise because our primary empirical approach relies on variation based on time and space differences. Third, most of the city-level implementation of the NRPS in this province occurred around 2013. Other cities (or areas within other provinces) had either already adopted the NRPS before 2013, or no information for their NRPS implementation at the city level was readily available online. Based on these public announcements, we can identify whether a city (or communities within a city) implemented the NRPS in a given year from 2009 to 2013, our primary analysis period. We can also identify when specific cities (and communities within these cities) switched from non-participation to participation in the NRPS.

[51] In this additional estimation exercise, we can estimate the original specifications at the city level. Specifically, we can compute the treatment intensity (percent of the city participating in the NRPS) based on the following formula: $city_participation_t$=(# communities in a city that offered the $NRPS_{ct}$) (# total communities in a city)$_{ct}$. This additional robustness check's main advantage is that we can observe the number of communities that implement the NRPS program based on public announcements (the numerator). The denominator of the fraction presented above is the total number of communities, and that number is a fixed constant. A significant disadvantage of this approach is that we can re-estimate the specifications from Section III only at the city level (and only for the Heilongjiang Province. We can obtain data on city or community announcements regarding NRPS implementation). This implies that in this additional analysis, the number of observations is low, limiting the statistical power for statistical inference.

However, we face a challenge relating to how the CHARLS defines the *community* unit differs from the administrative units available in public announcements. We cannot map the actual communities (within cities) to the community units (i.e., the variable *community ID*) in the CHARLS survey, the primary analysis unit in our analyses. This mismatch is due to an inconsistent definition of the "community" variable in the CHARLS and the administrative *community unit* (available in online records). Due to this mismatch, we cannot re-estimate the main specifications (performed at the community level), and therefore, we rely instead on the administrative data for community-level participation. However, we can re-estimate our primary specifications at the city level (i.e., at a higher geographical level than the community level used in the primary analysis). We execute this auxiliary exercise at the city level because: (1) we observe the actual number of communities, based on online public announcements, within a city that implemented the NRPS, and (2) we know the total number of communities, a fixed constant, in a city.

Therefore, we redefine our primary treatment variable for this additional empirical exercise: we change the treatment definition from the binary variable used in our primary analyses (at the community level) to a continuous variable that measures treatment intensity (at the city level). Based on this reconstructed treatment definition, we re-estimate this robustness check at the city level instead of the community level.

Using data for the Heilongjiang province, we re-estimate the main specifications outlined in Section III. However, we use a continuously defined treatment variable. We report this final robustness check at the city level in Online Appendix Table B6. This additional analysis (at the city level) relies on a minimal sample. Despite these shortcomings, the results echo the pattern reported earlier in Section IV. The effect size and direction of the program effects are consistent with our main estimates based on the CHARLS. Therefore, this additional analysis bolsters the results from the primary analysis.

## VI. Conclusion

This study examines the link between access to pension benefits and human capital depreciation in the form of cognitive decline among the elderly in rural China. Using the recently available longitudinal data from the CHARLS, we examine the effects of NRPS availability on two categories of cognitive functioning among the elderly: episodic memory and intact mental status.

Our analysis reveals a consistent story: access to program benefits reduced cognition performance among elderly beneficiaries. The estimated program impacts are similar to other negative findings in high-income countries, such as the U.S., England, and the European Union (Rohwedder and Willis 2010; Mazzonna and Peracchi 2012). People who live in places where the NRPS program is offered perform much worse on cognitive tests than people who live in areas where the program is not offered. We find substantially larger program impacts on the cognition measure that tests delayed word recall. Previous neurological research has documented the importance of this measure in flagging individuals who are more likely to experience an early onset of dementia. Furthermore, our findings support the *mental retirement* hypothesis that decreased mental activity atrophies cognitive skills. We show that retirement plans can contribute to cognitive decline in old age.

This study's findings have implications beyond China. To better understand how to balance the advantages of providing access to pension plans while reducing some of the adverse impacts that retirement-related programs can produce on cognitive performance in old age, more research is needed. Two specific areas deserve further exploration. First, what role does the type of job one has before retirement—formal versus informal or white-collar versus blue-collar—play in determining the speed of individual mental decline? It is crucial to shed greater insight into the underlying mechanisms that link pension benefits, retirement, and cognitive function.

Policy interventions targeting the elderly can have powerful economic consequences. Even mild cognitive deficits in the elderly result in a loss of quality of life and have detrimental effects on wellbeing. The results of this study point to a number of crucial areas where governments in LMICs could enhance individual welfare. Policies that aim to slow cognitive decline in older ages are likely to generate significant positive social spillovers. Our analysis does not offer guidance on which interventions are most effective but calls for a greater focus on policies aimed at facilitating social engagement and mental stimulation among the elderly.

# Figures and Tables

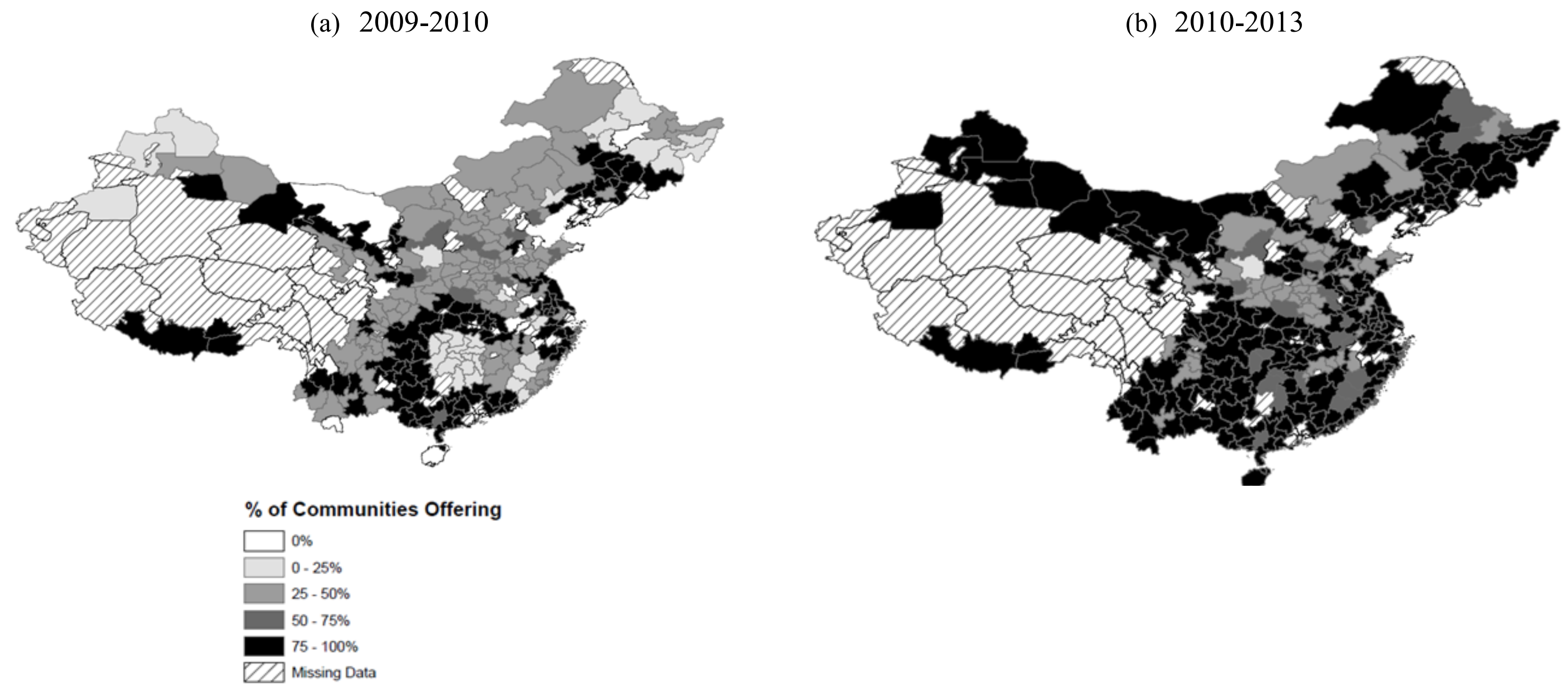

**Fig 1.** Geographic Implementation of NRPS. This figure shows the timely implementation of NRPS. "% of Communities Offering" indicates the percent of communities (*shequs*) within the province that implemented the NRPS.

**Table 1:** Summary Statistics.

| | Full Sample | Baseline | | |
|---|---|---|---|---|
| | | NRPS Participants | NRPS Non-Participants | p-value[a] |
| *Demographics of Respondents* | | | | |
| Respondent's Age | 59.31 (10.01) | 58.43 (9.68) | 58.44 (10.24) | 0.99 |
| # of Household Residents | 3.74 (1.87) | 3.68 (1.78) | 3.75 (1.88) | 0.04 |
| # Living Children | 2.77 (1.44) | 2.81 (1.39) | 2.74 (1.45) | 0.07 |
| Percent Female | 0.53 (0.50) | 0.54 (0.50) | 0.53 (0.50) | 0.38 |
| Percent Married | 0.80 (0.40) | 0.81 (0.39) | 0.78 (0.41) | 0.00 |
| Percent Living Near Children | 0.90 (0.30) | 0.91 (0.28) | 0.92 (0.27) | 0.40 |
| Percent With At Least Lower Secondary Education | 0.48 (0.50) | 0.48 (0.50) | 0.46 (0.50) | 0.10 |
| *Labour Market and Health Outcomes* | | | | |
| Weekly Work Hours | 45.45 (23.87) | 47.26 (24.07) | 46.89 (22.70) | 0.50 |
| Percent Currently Working | 0.70 (0.46) | 0.70 (0.46) | 0.69 (0.46) | 0.11 |
| Percent Working in Agriculture | 0.72 (0.45) | 0.72 (0.45) | 0.73 (0.45) | 0.49 |
| Percent Reporting Poor/Fair Health | 0.25 (0.43) | 0.27 (0.44) | 0.26 (0.44) | 0.23 |
| Respondent's BMI | 23.40 (3.84) | 23.62 (3.91) | 23.05 (3.81) | 0.00 |
| Percent Visited Doctor (Past Month) | 0.20 (0.40) | 0.20 (0.40) | 0.19 (0.39) | 0.08 |
| Percent Stayed in Hospital (Past Year) | 0.11 (0.31) | 0.10 (0.29) | 0.09 (0.28) | 0.06 |
| Percent Ever Smoked | 0.41 (0.49) | 0.40 (0.49) | 0.40 (0.49) | 0.98 |
| Percent Smoking Now | 0.25 (0.44) | 0.29 (0.45) | 0.30 (0.46) | 0.40 |
| *Cognition*[b] | | | | |
| Immediate Recall Score | 3.79 (1.76) | 3.93 (1.69) | 3.77 (1.70) | 0.00 |
| Delayed Recall Score | 2.86 (2.00) | 2.91 (1.91) | 2.89 (1.96) | 0.61 |
| Total Recall Score | 6.67 (3.47) | 6.85 (3.32) | 6.68 (3.36) | 0.02 |
| Cognitive Memory Index | 0.00 (1.43) | 0.06 (1.38) | 0.00 (1.39) | 0.06 |
| Observations | 28,034[c] | 10,011 | 3,680 | |

*Notes*: Standard deviations are reported in parenthesis. The full sample consists of observations from the 2011 and 2013 waves, whereas baseline observations are only for the sub-sample of participants and non-participants from the 2011 wave. (a) We test the null hypothesis that the difference in participant and non-participant means is equal to 0. (b) Low (or Negative) values denote lower performance on the cognition test. (c) Includes observations from 2011 and 2013 waves.

**Table 2:** NRPS Participation and Cognitive Performance.

| | Immediate Word Recall[a] | Immediate Word Recall[a] | Delay Word Recall[a] | Delay Word Recall[a] | Total Recall[a] | Total Recall[a] | Cognitive Memory Index[b] | Cognitive Memory Index[b] |
|---|---|---|---|---|---|---|---|---|
| | (1) | (2) | (3) | (4) | (5) | (6) | (7) | (8) |
| Panel A (ITT): | | | | | | | | |
| Offered NRPS × Above60 [c] | -0.092* | 0.134** | -0.187*** | -0.272*** | -0.278*** | -0.407*** | -0.093** | -0.123*** |
| | (0.053) | (0.056) | (0.054) | (0.063) | (0.094) | (0.107) | (0.038) | (0.041) |
| Baseline Mean | 3.792 | 3.792 | 2.862 | 2.862 | 6.678 | 6.678 | 0.000 | 0.000 |
| Controls | Yes | Yes | Yes | Yes | Yes | Yes | Yes | Yes |
| R-squared | 0.230 | 0.320 | 0.215 | 0.318 | 0.247 | 0.375 | 0.313 | 0.483 |
| Observations | 21,202 | 21,202 | 21,202 | 21,202 | 21,202 | 21,202 | 21,202 | 21,202 |
| Panel B (TOT): | | | | | | | | |
| NRPS Participation × Above60 [d] | -0.208* | -0.270** | -0.425*** | -0.547*** | -0.633*** | -0.816*** | -0.212** | -0.247*** |
| | (0.120) | (0.113) | (0.122) | (0.129) | (0.214) | (0.217) | (0.087) | (0.084) |
| Baseline Mean | 3.792 | 3.792 | 2.862 | 2.862 | 6.678 | 6.678 | 0.000 | 0.000 |
| Controls | Yes | Yes | Yes | Yes | Yes | Yes | Yes | Yes |
| Individual FE | No | Yes | No | Yes | No | Yes | No | Yes |
| F-Stat (First Stage) | 367.70 | 436.96 | 367.70 | 436.96 | 367.70 | 436.96 | 367.70 | 436.96 |
| R-squared | 0.065 | 0.003 | 0.152 | 0.004 | 0.102 | 0.005 | 0.110 | 0.004 |
| Observations | 21,202 | 21,202 | 21,202 | 21,202 | 21,202 | 21,202 | 21,202 | 21,202 |

*Notes*: The table reports estimates of the DDD estimator for the NRPS treatment effect. (a) Word recall tests: Immediate Recall = [0,10], Delayed Recall = [0,10] and Total Recall = [0,20]. (b) We created the Cognitive Memory Index using principal component analysis, combing short/long-term memory measures, working memory (Serial -7 Test) and orientation (Knowing the Current Month), and self-rated memory. (c) Our DDD coefficient: NRPS availability interacted with an indicator for being over 60 years old. (d) Individual participation is instrumented with the NRPS availability in the local municipality. Individual-level controls: Above60 (1= Yes), Education Levels (Base Group is illiterate with no formal education), # of Household Residents. Gender (=1 if Female) and Marital Status (=1 if Married) are included in the specifications without FEs. Even number columns include individual FE. Panel A is estimated using Ordinary Least Squares (OLS) with Community, Year, and Community×Year FE. Panel B is estimated using Two-Stage Least Squares (2SLS) with Community, Year, and Community×Year FE. Clustered standard errors at the community level are reported in parenthesis.
***Significant at the 1 percent level.
**Significant at the 5 percent level.
*Significant at the 10 percent level.

**Table 3**: Actual Retirement and Cognitive Performance.

| | Immediate Word Recall[b] | Immediate Word Recall[b] | Delay Word Recall[b] | Delay Word Recall[b] | Total Recall[b] | Total Recall[b] | Cognitive Memory Index[c] | Cognitive Memory Index[c] |
|---|---|---|---|---|---|---|---|---|
| | (1) | (2) | (3) | (4) | (5) | (6) | (7) | (8) |
| Retired | -0.469* | -0.810*** | -0.575 | 0.703 | -0.906 | -0.107 | -0.529** | -0.556 |
| (Yes=1)[a] | (0.283) | (0.189) | (0.369) | (1.478) | (0.613) | (1.570) | (0.252) | (1.220) |
| Baseline Mean | 3.792 | 3.792 | 2.862 | 2.862 | 6.678 | 6.678 | 0.000 | 0.000 |
| Controls | Yes | Yes | Yes | Yes | Yes | Yes | Yes | Yes |
| Individual FE | No | Yes | No | Yes | No | Yes | No | Yes |
| R-squared | 0.081 | 0.010 | 0.076 | 0.002 | 0.100 | 0.009 | 0.166 | 0.004 |
| F-Stat (First Stage) | 47.48 | 23.97 | 48.18 | 23.97 | 47.39 | 23.97 | 45.82 | 23.97 |
| Observations | 22,444 | 22,444 | 22,329 | 22,329 | 22,226 | 22,226 | 21,258 | 21,258 |

*Notes*: The table reports estimates of the DDD estimator for the NRPS treatment effect. (a) Directly asked about retirement procedure. "Have you completed retirement procedures (including early retirement) or internal retirement (Retirement from government departments, enterprises, and institutions, not including retirement in the sense of getting agricultural insurance)?" A positive answer is coded as being retired. We also use data from additional variables: the person completed retirement procedures in any survey wave, the reported number of days (or months, hours) worked is zero in three consecutive waves, and the reported usual number of days (or months) per year is zero for three consecutive waves, the reported monetary retirement benefit is positive, the number of workdays missed for health reasons has been more than 300 per year for three consecutive survey waves, reported year of retirement is before 2009, and the survey respondent indicated that the formal retirement is processed. (b) Word recall tests: Immediate Recall = [0,10], Delayed Recall = [0,10] and Total Recall = [0,20]. (c) We created the Cognitive Memory Index using principal component analysis, combing short/long-term memory measures, working memory (Serial -7 Test) and orientation (Knowing the Current Month), and self-rated memory. Individual level controls: Above60 (1= Yes), Education Levels (Base Group is illiterate with no formal education), # of Household Residents. Gender (=1 if Female) and Marital Status (=1 if Married) are only included in the specifications without FEs. Even number columns include individual FE and Community, Year, and Community×Year FE. Clustered standard errors at the community level are reported in parenthesis. The number of observations in this table differs from Table 2 because of the different independent variables used.
***Significant at the 1 percent level.
**Significant at the 5 percent level.
*Significant at the 10 percent level

**Table 4**: Duration of NRPS Benefits and Cognitive Decline.

| | Immediate Word Recall[b] | Immediate Word Recall[b] | Delay Word Recall[b] | Delay Word Recall[b] | Total Recall[b] | Total Recall[b] | Cognitive Memory Index[c] | Cognitive Memory Index[c] |
|---|---|---|---|---|---|---|---|---|
| | (1) | (2) | (3) | (4) | (5) | (6) | (7) | (8) |
| Offer NRPS × Above 60 | -0.047 | -1.215 | 0.375 | 1.383 | 0.328 | 0.168 | -0.052 | -0.728 |
| | (0.487) | (1.885) | (0.482) | (1.891) | (0.874) | (3.632) | (0.378) | (1.243) |
| Duration of NRPS Benefits | -0.154 | -0.584* | -0.122 | -0.629 | -0.276 | -1.213* | -0.088 | -0.647** |
| (Receiving for 1-2 years) | (0.144) | (0.308) | (0.158) | (0.425) | (0.264) | (0.613) | (0.109) | (0.290) |
| Duration of NRPS Benefits | 0.204 | -0.619 | 0.503 | -0.186 | 0.707 | -0.805 | 0.333 | -0.603 |
| (Receiving for 3 or more years) | (0.316) | (0.602) | (0.364) | (0.886) | (0.534) | (1.350) | (0.225) | (0.548) |
| Offer NRPS (Yes=1) × Above 60 (Yes=1) | 0.168 | 1.490 | -0.046 | 0.212 | 0.122 | 1.703 | 0.111 | 1.292* |
| × Duration of NRPS Benefits (1-2 years)[a] | (0.186) | (0.919) | (0.210) | (0.928) | (0.344) | (1.531) | (0.143) | (0.664) |
| Offer NRPS (Yes=1) × Above 60 (Yes=1) × Duration of NRPS Benefits (3 years or | -0.325 | 1.674 | -0.727* | 0.070 | -1.052* | 1.744 | -0.382 | 1.445* |
| more)[a] | (0.339) | (1.014) | (0.393) | (1.206) | (0.582) | (1.894) | (0.246) | (0.808) |
| Above 60 | -0.280 | -0.398 | -0.273 | -1.135 | -0.554 | -1.533 | -0.193 | -0.393 |
| | (0.419) | (0.998) | (0.382) | (1.326) | (0.725) | (2.229) | (0.322) | (1.092) |
| Baseline Mean | 3.792 | 3.792 | 2.862 | 2.862 | 6.678 | 6.678 | 0.000 | 0.000 |
| Controls | Yes | Yes | Yes | Yes | Yes | Yes | Yes | Yes |
| Individual FE | No | Yes | No | Yes | No | Yes | No | Yes |
| R-squared | 0.234 | 0.781 | 0.225 | 0.760 | 0.252 | 0.786 | 0.310 | 0.818 |
| Observations | 22,199 | 22,199 | 22,092 | 22,092 | 21,992 | 21,992 | 21,041 | 21,041 |

*Notes*: The table reports estimates of the DDD estimator for the NRPS treatment effect. (a) Directly asked about retirement procedure. "Have you completed retirement procedures (including early retirement) or internal retirement (Retirement from government departments, enterprises, and institutions, not including retirement in the sense of getting agricultural insurance)?" A positive answer is coded as being retired. (b) Word recall tests: Immediate Recall = [0,10], Delayed Recall = [0,10] and Total Recall = [0,20]. (c) We created the Cognitive Memory Index using principal component analysis, combing short/long-term memory measures, working memory (Serial -7 Test) and orientation (Knowing the Current Month), and self-rated memory. Individual level controls: Above60 (1= Yes), Education Levels (Base Group is illiterate with no formal education), # of Household Residents. Gender (=1 if Female) and Marital Status (=1 if Married) are only included in the specifications without FEs. Even number columns include individual fixed effects and Community, Year, and Community×Year FE. Clustered standard errors at the community level are reported in parenthesis. The number of observations in this table differs from Table 2 because of the different independent variables used.

***Significant at the 1 percent level.

**Significant at the 5 percent level.

*Significant at the 10 percent level

**Table 5**: Mechanisms Analysis (Labor, Mental and Social Engagements)

| | Labor Activities | | | | Mental Stimulation | | | Social Engagement | | | | |
|---|---|---|---|---|---|---|---|---|---|---|---|---|
| | #Months Worked (Past year) | Hours Daily Worked (Per Week) | Self-Employment #Months Worked (Past year) | Self-Employment Hours Daily Worked (Per Week) | Played Majong Last Month (Yes=1) | Adult Education Course Last Month (Yes=1) | Mental Stimulation Index | Helped Friends Last Month (Yes=1) | Any Community Activity Last Month (Yes=1) | Volunteered Last Month (Yes=1) | Interact w Friends Last Month (Yes=1) | Social Engagement Index |
| | (1) | (2) | (3) | (4) | (5) | (6) | (7) | (8) | (9) | (10) | (11) | (12) |
| Panel A (ITT): | | | | | | | | | | | | |
| Offered NRPS × Above60 [a] | -0.160 | -0.024 | -1.342** | -1.224** | -0.011 | -0.002 | -0.043 | -0.023** | -0.026 | -0.040*** | -0.034*** | -0.105** |
| | (0.464) | (0.034) | (0.565) | (0.516) | (0.010) | (0.001) | (0.028) | (0.009) | (0.016) | (0.012) | (0.012) | (0.040) |
| Baseline Mean | 7.942 | 3.581 | 8.876 | 7.874 | 0.607 | 0.032 | 0.548 | 0.099 | 0.095 | 0.061 | 0.546 | 0.000 |
| Controls | Yes | Yes | Yes | Yes | Yes | Yes | Yes | Yes | Yes | Yes | Yes | Yes |
| R-squared | 0.381 | 0.154 | 0.379 | 0.373 | 0.441 | 0.264 | 0.350 | 0.146 | 0.316 | 0.180 | 0.356 | 0.115 |
| Observations | 2,668 | 15,203 | 1,748 | 1,701 | 12,842 | 11,134 | 11,818 | 21,198 | 21,198 | 21,198 | 21,198 | 21,198 |
| Panel B (TOT): | | | | | | | | | | | | |
| NRPS Participation × Above60 [b] | -0.377 | -0.055 | -3.151** | -2.968** | -0.025 | -0.003 | -0.099 | -0.052** | -0.050 | -0.063*** | -0.075*** | -0.239** |
| | (1.095) | (0.079) | (1.346) | (1.263) | (0.022) | (0.003) | (0.065) | (0.021) | (0.033) | (0.019) | (0.028) | (0.094) |
| Baseline Mean | 7.942 | 3.581 | 8.876 | 7.874 | 0.607 | 0.032 | 0.548 | 0.099 | 0.095 | 0.061 | 0.546 | 0.000 |
| Controls | Yes | Yes | Yes | Yes | Yes | Yes | Yes | Yes | Yes | Yes | Yes | Yes |
| F-Stat (First Stage) | 38.99 | 195.89 | 36.97 | 32.48 | 238.71 | 234.67 | 236.72 | 291.12 | 235.37 | 234.65 | 239.65 | 291.12 |
| R-squared | 0.380 | 0.154 | 0.357 | 0.357 | 0.034 | 0.012 | 0.030 | 0.012 | 0.004 | 0.006 | 0.030 | 0.001 |
| Observations | 2,668 | 15,203 | 1,748 | 1,701 | 12,842 | 11,134 | 11,818 | 21,198 | 11,178 | 11,145 | 14,775 | 21,198 |

*Notes*: The number of observations reflects the number of non-missing values for each outcome in the CHARLS. The table reports estimates of the DDD estimator for the NRPS treatment effect. Food expenses are in constant 2011 Yuan. (a) The DDD estimator (NRPS availability interacted with an indicator for being over 60 years old). The control group is individuals under 60 living in eligible communities that did not offer NRPS between 2011 and 2013. (b) Individual participation is instrumented with the policy variable. Individual-level controls: Above60 (1= Yes), Education Levels (Base Group is illiterate with no formal education), # of Household Residents. Gender (=1 if Female) and Marital Status (=1 if Married) are only included in the specifications without FEs. Regular alcohol drinker: drinking at least once per week in the last year. Panel A is estimated using Ordinary Least Squares (OLS) with Community, Year, and Community*Year FE. Panel B is estimated using Two-Stage Least Squares (2SLS) with Community, Year, and Community×Year FE. Clustered standard errors at the community level are reported in parenthesis. In constant 2011 Yuan. *p< 0.10, **p< 0.05, ***p< 0.01.

**Table 6**: Mechanisms Analysis (Health Behaviors).

| | Health Behaviors and Nutrition | | | | |
|---|---|---|---|---|---|
| | Hrs Sleep per Night (Last Year) | Currently Smoking (Yes=1) | Regular Alcohol Drinker (Yes=1) | Health Behaviors Index | HH Food Expenses (Last week in Yuan) |
| | (1) | (2) | (3) | (4) | (5) |
| Panel A (ITT): | | | | | |
| Offered NRPS × Above60 [a] | 0.157** | -0.019* | -0.020** | 0.083** | 17.434 |
| | (0.061) | (0.011) | (0.009) | (0.035) | (39.528) |
| Baseline Mean | 6.281 | 0.254 | 0.186 | 0.000 | 192.443 |
| Controls | Yes | Yes | Yes | Yes | Yes |
| R-squared | 0.107 | 0.389 | 0.257 | 0.114 | 0.052 |
| Observations | 20,913 | 18,965 | 20,688 | 18,199 | 23,822 |
| Panel B (TOT): | | | | | |
| NRPS Participation × Above60 [b] | 0.354** | -0.042* | -0.045** | 0.185** | 37.767 |
| | (0.142) | (0.025) | (0.020) | (0.079) | (85.542) |
| Baseline Mean | 6.281 | 0.254 | 0.186 | 0.000 | 178.488 |
| Controls | Yes | Yes | Yes | Yes | Yes |
| F-Stat (First Stage) | 244.47 | 223.79 | 243.02 | 296.37 | 244.743 |
| R-squared | 0.104 | 0.389 | 0.256 | 0.012 | 0.053 |
| Observations | 20,913 | 18,965 | 20,688 | 18,199 | 23,822 |

*Notes*: The number of observations reflects the number of non-missing values for each outcome in the CHARLS. The table reports estimates of the DDD estimator for the NRPS treatment effect. Food expenses are in constant 2011 Yuan. (a) The DDD estimator (NRPS availability interacted with an indicator for being over 60 years old). The control group is individuals under 60 living in eligible communities that did not offer NRPS between 2011 and 2013. (b) Individual participation is instrumented with the policy variable. Individual-level controls: Above60 (1= Yes), Education Levels (Base Group is illiterate with no formal education), # of Household Residents. Gender (=1 if Female) and Marital Status (=1 if Married) are included in the specifications without FEs. Regular alcohol drinker: drinking at least once per week in the last year. Panel A is estimated using Ordinary Least Squares (OLS) with Community, Year, and Community*Year FE. Panel B is estimated using Two-Stage Least Squares (2SLS) with Community, Year, and Community×Year FE. Clustered standard errors at the community level are reported in parenthesis. In constant 2011 Yuan. *p< 0.10, **p< 0.05, ***p< 0.01.

# Appendix A

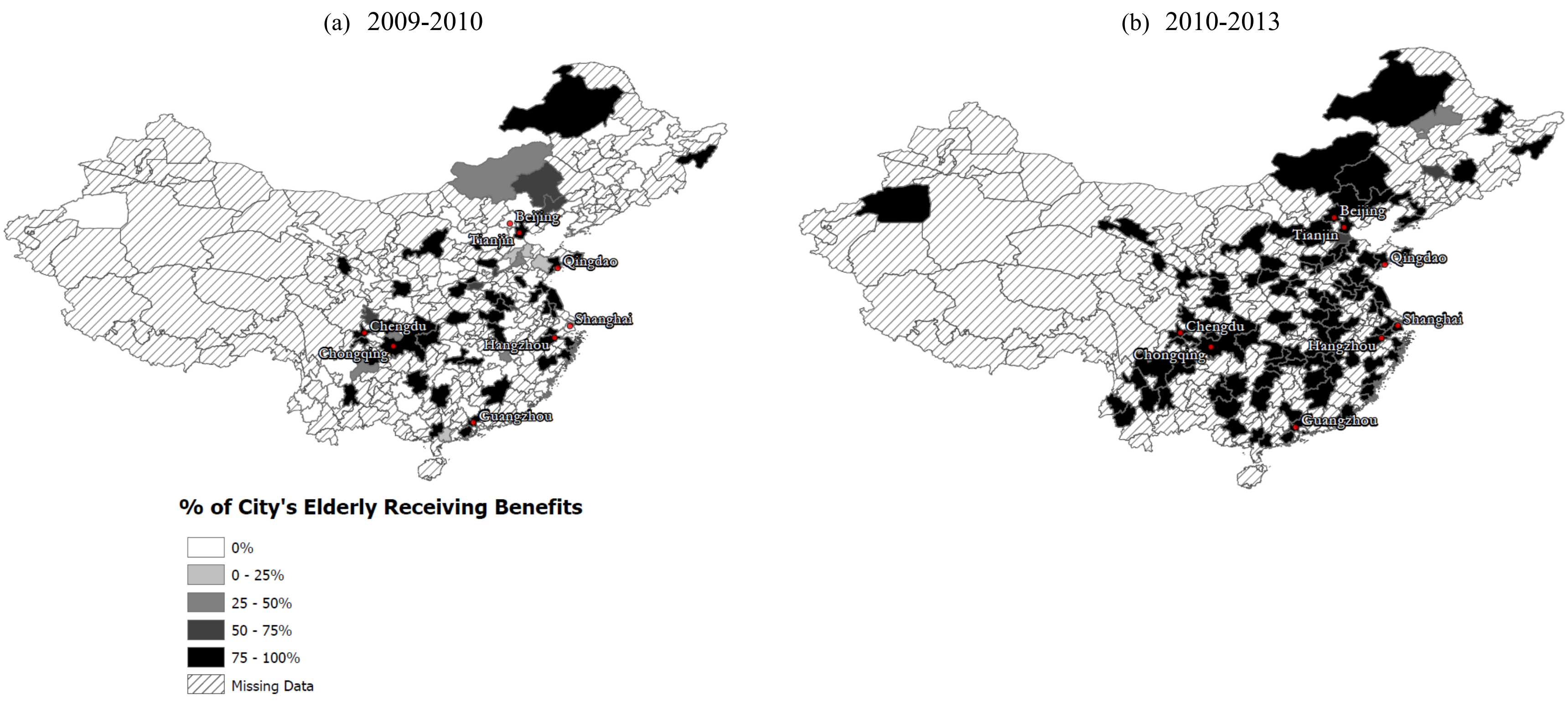

**Fig A1**. Geographic Implementation of NRPS. This figure reports the percentage of the elderly receiving NRPS benefits as a fraction of all elderly in each community. “% of City’s Elderly Receiving Benefits Offering” = the number of elderly receiving NRPS benefits in a community (*shequ*)/the number of elderly in a community.

**Table A1.** Summary Statistics in NRPS Implementing and Non-implementing Areas.

| | 2011 | | | 2013 | | |
|---|---|---|---|---|---|---|
| | Implementing | Non-implementing | p-value | Implementing | Non-implementing | p-value |
| *Demographics of Respondents* | | | | | | |
| Han Nationality | 0.935 (0.202) | 0.915 (0.217) | 0.344 | 0.925 (0.208) | 0.933 (0.233) | 0.849 |
| Respondent's Age | 58.111 (3.275) | 58.194 (5.129) | 0.841 | 60.347 (3.749) | 55.924 (9.743) | 0.001 |
| # of Household Residents | 3.649 (0.831) | 3.811 (1.013) | 0.072 | 3.789 (0.918) | 2.903 (0.979) | 0.001 |
| # Living Children | 2.685 (0.611) | 2.641 (0.756) | 0.510 | 2.733 (0.664) | 2.246 (1.24) | 0.001 |
| Percent Female | 0.535 (0.076) | 0.556 (0.157) | 0.070 | 0.541 (0.084) | 0.591 (0.332) | 0.025 |
| Percent Married | 0.785 (0.16) | 0.734 (0.24) | 0.011 | 0.788 (0.155) | 0.837 (0.306) | 0.161 |
| Percent With At Least Lower Secondary Education | 0.446 (0.173) | 0.444 (0.224) | 0.929 | 0.433 (0.176) | 0.601 (0.359) | 0.001 |
| *Labor Market and Health Outcomes* | | | | | | |
| Weekly Work Hours | 3.702 (0.389) | 3.634 (0.539) | 0.149 | 3.513 (0.482) | 4.005 (0.406) | 0.001 |
| Percent Currently Working | 0.66 (0.191) | 0.615 (0.258) | 0.040 | 0.682 (0.183) | 0.569 (0.445) | 0.009 |
| Percent Working in Agriculture | 0.592 (0.317) | 0.612 (0.356) | 0.567 | 0.609 (0.313) | 0.102 (0.256) | 0.001 |
| Percent Reporting Poor/Fair Health | 0.303 (0.18) | 0.267 (0.237) | 0.083 | 0.263 (0.161) | 0.352 (0.334) | 0.015 |
| Respondent's BMI | 23.907 (2.516) | 23.127 (1.902) | 0.001 | 23.771 (1.702) | 26.948 (5.756) | 0.001 |
| Percent Visited Doctor (Past Month) | 0.186 (0.121) | 0.187 (0.181) | 0.958 | 0.205 (0.139) | 0.193 (0.342) | 0.706 |
| Percent Stayed in Hospital (Past Year) | 0.085 (0.066) | 0.073 (0.086) | 0.133 | 0.124 (0.084) | 0.181 (0.265) | 0.009 |
| Percent Ever Smoked | 0.286 (0.134) | 0.284 (0.174) | 0.883 | 0.182 (0.134) | 0.179 (0.282) | 0.923 |
| Immediate Recall Score | 3.905 (0.747) | 3.892 (0.885) | 0.878 | 3.859 (0.736) | 3.71 (1.468) | 0.370 |
| Delayed Recall Score | 2.918 (0.765) | 2.942 (0.969) | 0.778 | 2.944 (0.819) | 2.446 (1.861) | 0.009 |
| Total Recall Score | 6.834 (1.416) | 6.846 (1.721) | 0.938 | 6.832 (1.482) | 6.309 (2.874) | 0.118 |
| Cognitive Memory Index | 0.084 (0.604) | 0.105 (0.72) | 0.740 | 0.057 (0.585) | 0.077 (1.234) | 0.882 |
| Observation (Communities) | 219 | 195 | | 394 | 24 | |

*Notes*: (a) Standard deviations are reported in parenthesis. (b) We test the null hypothesis that the difference between participating community and non-participating community means is equal to 0.

**Table A2:** Test of Common Trends Using CHNS Data.

| | | Immediate Word Recall[a] | Delayed Word Recall[a] | Total Recall[a] |
|---|---|---|---|---|
| | | (1) | (2) | (3) |
| 50% Coverage Rate Threshold | Treatment × Age>Above 60 (Yes=1) × 2000 | -0.216 | -0.286 | -0.808 |
| | | (0.387) | (0.399) | (0.761) |
| | Treatment × Age>Above 60 (Yes=1) × 2006 | -0.446 | -0.325 | -0.708 |
| | | (0.306) | (0.383) | (0.586) |
| | R-Squared Adj | 0.215 | 0.228 | 0.241 |
| | Year FE | Yes | Yes | Yes |
| | Community FE | Yes | Yes | Yes |
| | Observations | 4,742 | 4,719 | 4,615 |
| 70% Coverage Rate Threshold | Treatment × Age>Above 60 (Yes=1) × 2000 | -0.186 | 0.519 | 0.364 |
| | | (0.417) | (0.465) | (0.837) |
| | Treatment × Age>Above 60 (Yes=1) × 2006 | -0.479 | 0.173 | -0.281 |
| | | (0.309) | (0.383) | (0.660) |
| | R-Squared Adj | 0.214 | 0.229 | 0.241 |
| | Year FE | Yes | Yes | Yes |
| | Community FE | Yes | Yes | Yes |
| | Observations | 4,742 | 4,719 | 4,615 |

*Notes*: Source: CHNS 2000, 2004, and 2006 Waves. The base year is 2004. (a) Word recall tests: Immediate Recall = [0,10], Delayed Recall = [0,10] and Total Recall = [0,20]. All columns include Community and Year FE. Clustered standard errors at the community level are reported in parenthesis.

***Significant at the 1 percent level.

**Significant at the 5 percent level.

*Significant at the 10 percent level

**Table A3:** Test of Common Trends Using CHNS Data (Threshold Rate Scenarios).

| | | Immediate Word Recall[a] | Delayed Word Recall[a] | Total Recall[a] |
|---|---|---|---|---|
| | | (1) | (2) | (3) |
| 20% Coverage Rate Threshold | Treatment × Age>Above 60 | 0.232 | 0.389 | 0.631 |
| | (Yes=1) × 2000 | (0.469) | (0.490) | (0.934) |
| | Treatment × Age>Above 60 | 0.070 | 0.235 | 0.287 |
| | (Yes=1) × 2006 | (0.517) | (0.437) | (0.867) |
| | R-Squared Adj | 0.214 | 0.228 | 0.242 |
| | Year FE | Yes | Yes | Yes |
| | Community FE | Yes | Yes | Yes |
| | Observations | 4,711 | 4,689 | 4,584 |
| 30% Coverage Rate Threshold | Treatment × Age>Above 60 | 0.308 | 0.473 | 0.845 |
| | (Yes=1) × 2000 | (0.385) | (0.413) | (0.770) |
| | Treatment × Age>Above 60 | 0.027 | 0.184 | 0.206 |
| | (Yes=1) × 2006 | (0.330) | (0.338) | (0.624) |
| | R-Squared Adj | 0.215 | 0.229 | 0.242 |
| | Year FE | Yes | Yes | Yes |
| | Community FE | Yes | Yes | Yes |
| | Observations | 4,711 | 4,689 | 4,584 |
| 50% Coverage Rate Threshold | Treatment × Age>Above 60 | 0.435 | 0.557 | 0.940 |
| | (Yes=1) × 2000 | (0.378) | (0.386) | (0.735) |
| | Treatment × Age>Above 60 | 0.128 | 0.248 | 0.315 |
| | (Yes=1) × 2006 | (0.313) | (0.324) | (0.596) |
| | R-Squared Adj | 0.215 | 0.229 | 0.242 |
| | Year FE | Yes | Yes | Yes |
| | Community FE | Yes | Yes | Yes |
| | Observations | 4,711 | 4,689 | 4,584 |
| 62% Coverage Rate Threshold | Treatment × Age>Above 60 | 0.535 | 0.702 | 1.167 |
| | (Yes=1) × 2000 | (0.501) | (0.469) | (0.935) |
| | Treatment × Age>Above 60 | 0.092 | 0.474 | 0.490 |
| | (Yes=1) × 2006 | (0.374) | (0.385) | (0.697) |
| | R-Squared Adj | 0.215 | 0.229 | 0.242 |
| | Year FE | Yes | Yes | Yes |
| | Community FE | Yes | Yes | Yes |
| | Observations | 4,711 | 4,689 | 4,584 |

**Table A4**: NRPS Participation and Migration

| | Migrated between 2011 and 2013 (=1 if yes) |
|---|---|
| **Panel A: ITT** | |
| Offered NRPS × Above60 | 0.029 |
| | (0.128) |
| Mean of outcome | 0.020 |
| R-Squared | 0.261 |
| Control | Yes |
| Province FE | Yes |
| Observations | 11,763 |
| **Panel B: TOT** | |
| NRPS Participation × Above60 | 0.036 |
| | (0.174) |
| Mean of outcome | 0.020 |
| F-stat | 2,011.65 |
| Control | Yes |
| Province FE | Yes |
| Observations | 11,746 |

*Notes*: (a) The table reports estimates of the DDD estimator for the NRPS treatment effect. (b) Our DDD coefficient: NRPS availability interacted with an indicator for being over 60 years old. (c) Individual participation is instrumented with the NRPS availability in the local municipality. (d) Individual level controls: Above60 (1= Yes), Education Levels (Base Group is illiterate with no formal education), # of Household Residents. Gender (=1 if Female) and Marital Status (=1 if Married) are included. (e) Panel A is estimated using Ordinary Least Squares (OLS) with Community, Year, and Community×Year FE. Panel B is estimated using Two-Stage Least Squares (2SLS) with Community, Year, and Community×Year FE. (f) Clustered standard errors at the community level are reported in parenthesis.
***Significant at the 1 percent level.
**Significant at the 5 percent level.
*Significant at the 10 percent level.

**Table A5**: NRPS Participation and Attrition

| | Attrition between 2011 and 2013 (=1 if yes) |
|---|---|
| **Panel A: ITT** | |
| Offered NRPS × Above60 | 0.018 |
| | (0.100) |
| Mean of outcome | 0.166 |
| R-Squared | 0.031 |
| Control | Yes |
| Observations | 12,300 |
| **Panel B: TOT** | |
| NRPS Participation × Above60 | 0.011 |
| | (0.182) |
| Mean of outcome | 0.166 |
| F-stat | 2,029.55 |
| Control | Yes |
| Observations | 12,300 |

*Notes*: (a) The table reports estimates of the OLS and IV estimators for the NRPS treatment effect. (b) Individual participation is instrumented with the NRPS availability in the local municipality. (c) Individual level controls: Above60 (1= Yes), Education Levels (Base Group is illiterate with no formal education), # of Household Residents. Gender (=1 if Female) and Marital Status (=1 if Married) are included. (d) Panel A is estimated using Ordinary Least Squares (OLS), and Panel B is estimated using Two-Stage Least Squares (2SLS). (e) Clustered standard errors at the community level are reported in parenthesis.
***Significant at the 1 percent level.
**Significant at the 5 percent level.
*Significant at the 10 percent level.

**Table A6:** Heterogeneous Treatment Effects by Gender.

| | Immediate Word Recall[a] | Delayed Word Recall[a] | Total Recall[a] | Cognitive Index[b] |
|---|---|---|---|---|
| | (1) | (2) | (3) | (4) |
| Panel A (ITT): | | | | |
| Offered NRPS × Above60 × Female [c] | -0.092 | 0.073 | -0.019 | -0.008 |
| | (0.102) | (0.125) | (0.195) | (0.075) |
| Offered NRPS × Above60 | -0.036 | -0.249** | -0.285* | -0.070 |
| | (0.076) | (0.086) | (0.141) | (0.055) |
| Offered NRPS × Female | -0.005 | -0.100 | -0.105 | -0.025 |
| | (0.065) | (0.085) | (0.127) | (0.047) |
| Above60 × Female | 0.037 | -0.082 | -0.045 | -0.104* |
| | (0.079) | (0.092) | (0.150) | (0.061) |
| Above60 | -0.429*** | -0.407*** | -0.836*** | -0.268*** |
| | (0.058) | (0.066) | (0.110) | (0.045) |
| Female | 0.126** | 0.276*** | 0.402*** | 0.025 |
| | (0.050) | (0.059) | (0.091) | (0.035) |
| Baseline Mean | 3.792 | 2.862 | 6.678 | 0.000 |
| Controls | Yes | Yes | Yes | Yes |
| R-squared | 0.223 | 0.217 | 0.244 | 0.304 |
| Observations | 21,202 | 21,202 | 21,202 | 21,202 |
| Panel B (TOT): | | | | |
| NRPS Participation × Above60 × Female[d] | -0.209 | 0.152 | -0.057 | -0.021 |
| | (0.228) | (0.280) | (0.436) | (0.167) |
| NRPS Participation × Above60[d] | -0.076 | -0.452*** | -0.528** | -0.147 |
| | (0.168) | (0.176) | (0.305) | (0.118) |
| NRPS Participation × Female | -0.004 | -0.097 | -0.101 | -0.024 |
| | (0.065) | (0.085) | (0.127) | (0.047) |
| Above60 × Female | 0.081 | -0.101 | -0.020 | -0.097 |
| | (0.112) | (0.135) | (0.212) | (0.085) |
| Above60 | -0.415*** | -0.308*** | -0.723*** | -0.240*** |
| | (0.081) | (0.094) | (0.153) | (0.061) |
| Female | 0.124** | 0.273*** | 0.397*** | 0.024 |
| | (0.050) | (0.059) | (0.091) | (0.035) |
| Baseline Mean | 3.792 | 2.862 | 6.678 | 0.000 |
| Controls | Yes | Yes | Yes | Yes |
| R-squared | 0.110 | 0.100 | 0.124 | 0.193 |
| F-Stat (First Stage) | 141.13 | 141.13 | 141.13 | 141.13 |
| Observations | 21,202 | 21,202 | 21,202 | 21,202 |

*Notes:* The table reports estimates of the DDDD estimator for the NRPS treatment effect. (a) Word recall tests: Immediate Recall = [0,10], Delayed Recall = [0,10] and Total Recall = [0,20]. (b) We created the Cognitive Memory Index using principal component analysis, combing short/long-term memory measures, working memory (Serial -7 Test) and orientation (Knowing the Current Month), and self-rated memory. (c) DDD coefficient: NRPS availability interacted with an indicator for being over 60 years old. (d) Individual participation is instrumented with the NRPS availability in the local municipality. Individual-level controls: Age, Age Squared, Marital Status (=1 if Married), Gender (=1 if Female), Education Levels (Base Group is illiterate with no formal education), # of Household Residents. The ITT effects are estimated using Ordinary Least Squares (OLS) with Community, Year, and Community*Year FE. Panel B is estimated using Two-Stage Least Squares (2SLS) with Community, Year, and Community×Year FE. Clustered standard errors at the community level are reported in parenthesis.

***Significant at the 1 percent level.

**Significant at the 5 percent level.

*Significant at the 10 percent level

**Table A7**: Heterogeneous Treatment Effects by Household Members

| | Immediate Word Recall | | | Delay Word Recall | | | Total Recall | | | Cognitive Memory Index | | |
|---|---|---|---|---|---|---|---|---|---|---|---|---|
| | Up to 2 Members | 3-5 Members | Above 5 Members | Up to 2 Members | 3-5 Members | Above 5 Members | Up to 2 Members | 3-5 Members | Above 5 Members | Up to 2 Members | 3-5 Members | Above 5 Members |
| Panel A: ITT | | | | | | | | | | | | |
| Offered NRPS × Above60 | -0.167** | -0.089 | -0.007 | -0.276*** | -0.133 | -0.196 | -0.443*** | -0.223 | -0.203 | -0.123* | -0.058 | -0.082 |
| | (0.084) | (0.094) | (0.116) | (0.098) | (0.093) | (0.141) | (0.162) | (0.153) | (0.237) | (0.065) | (0.064) | (0.090) |
| Controls | Yes | Yes | Yes | Yes | Yes | Yes | Yes | Yes | Yes | Yes | Yes | Yes |
| R-squared | 0.293 | 0.261 | 0.331 | 0.267 | 0.260 | 0.313 | 0.302 | 0.285 | 0.340 | 0.363 | 0.345 | 0.402 |
| Observations | 6,540 | 9,555 | 3,282 | 6,540 | 9,555 | 3,282 | 6,540 | 9,555 | 3,282 | 6,540 | 9,555 | 3,282 |
| Panel B: TOT | | | | | | | | | | | | |
| Offered Participation × Above60 | -0.362** | -0.210 | -0.016 | -0.596*** | -0.313 | -0.425 | -0.958*** | -0.523 | -0.441 | -0.266* | -0.135 | -0.178 |
| | (0.181) | (0.221) | (0.251) | (0.217) | (0.221) | (0.309) | (0.354) | (0.361) | (0.515) | (0.141) | (0.150) | (0.194) |
| Controls | Yes | Yes | Yes | Yes | Yes | Yes | Yes | Yes | Yes | Yes | Yes | Yes |
| F-Stat (First Stage) | 274.04 | 239.07 | 94.68 | 274.04 | 239.07 | 94.68 | 274.04 | 239.07 | 94.68 | 274.04 | 239.07 | 94.68 |
| Observations | 6,540 | 9,555 | 3,282 | 6,540 | 9,555 | 3,282 | 6,540 | 9,555 | 3,282 | 6,540 | 9,555 | 3,282 |

*Notes*: The table reports estimates of the DDD estimator for the NRPS treatment effect by household members. (a) Word recall tests: Immediate Recall = [0,10], Delayed Recall = [0,10] and Total Recall = [0,20]. (b) We created the Cognitive Memory Index using principal component analysis, combing measures of short/long term memory, working memory (Serial -7 Test) and orientation (Knowing the Current Month), and self-rated memory. (c) DDD coefficient: NRPS availability interacted with an indicator for being over 60 years old. (d) Individual participation is instrumented with the NRPS availability in the local municipality. Individual level controls: Age, Age Squared, Marital Status (=1 if Married), Gender (=1 if Female), and Education Levels (Base Group is illiterate with no formal education). The ITT effects are estimated using Ordinary Least Squares (OLS) with Community, Year, and Community×Year FE. Panel B is estimated using Two-Stage Least Squares (2SLS) with Community, Year, and Community×Year FE. Clustered standard errors at the community level are reported in parenthesis.
***Significant at the 1 percent level.
**Significant at the 5 percent level.
*Significant at the 10 percent level.

**Table A8**: Falsification Test Using Placebo Sample.

| | Immediate Word Recall[a] | Total Recall[a] | Cognitive Memory Index[b] |
|---|---|---|---|
| | (1) | (2) | (3) |
| Offered NRPS × Above60[c] | -0.098 | -0.710 | -0.183 |
| | (0.245) | (0.494) | (0.207) |
| Baseline Mean | 0.253 | 0.000 | 0.000 |
| Controls | Yes | Yes | Yes |
| R-squared | 0.611 | 0.625 | 0.620 |
| Observations | 604 | 594 | 576 |

*Notes:* The table reports estimates of the DDD estimator for the NRPS treatment effect. (a) Word recall tests: Immediate Recall = [0,10], Delayed Recall = [0,10] and Total Recall = [0,20]. (b) We created the Cognitive Memory Index using principal component analysis, combing short/long-term memory measures, working memory (Serial -7 Test) and orientation (Knowing the Current Month), and self-rated memory. (c) Our DDD coefficient: NRPS availability interacted with an indicator for being over 60 years old. A significant coefficient suggests the differential treatment of urban pensioners in treated communities relative to urban pensioners in control communities; a cause of concern for the instrument's validity. Individual-level controls: Above60 (1=Yes), Marital Status (=1 if Married), Gender (=1 if Female), Education Levels (Base Group is illiterate with no formal education), # of Household Residents. The specifications are estimated with Community, Year, and Community×Year FE. Clustered standard errors at the community level are reported in parenthesis.
***Significant at the 1 percent level.
**Significant at the 5 percent level.
*Significant at the 10 percent level

**Table A9:** Test on Placebo Outcomes for Specifications (1) and (3).

| | Han (=1 if yes) | # Dead Daughter | Mother's Education | # of Living Sons |
|---|---|---|---|---|
| | (1) | (2) | (3) | (4) |
| Panel A (ITT): | | | | |
| Offered NRPS × Above60 [a] | -0.004 | -0.013 | -0.015 | 0.010 |
| | (0.004) | (0.029) | (0.012) | (0.024) |
| Baseline Mean | 0.920 | 1.299 | 1.190 | 1.466 |
| Controls | Yes | Yes | Yes | |
| R-squared | 0.652 | 0.165 | 0.130 | 0.235 |
| Observations | 20,102 | 21,202 | 19,656 | 21,202 |
| Panel B (TOT): | | | | |
| NRPS Participation × Above60 [b] | -0.010 | -0.032 | -0.035 | 0.025 |
| | (0.009) | (0.071) | (0.030) | (0.059) |
| Baseline Mean | 0.920 | 1.299 | 1.190 | 1.466 |
| Controls | Yes | Yes | Yes | Yes |
| F-Stat (First Stage) | 282.617 | 279.6213 | 291.9617 | 279.6213 |
| R-squared | 0.652 | 0.165 | 0.130 | 0.235 |
| Observations | 20,102 | 21,202 | 19,656 | 21,202 |

*Notes:* The table reports estimates of the DDD estimator for the NRPS treatment effect. (a) DDD coefficient(NRPS availability interacted with an indicator for being over 60 years old). The control group becomes individuals under 60 living in eligible communities that did not offer NRPS between 2011 and 2013. (b) Individual participation is instrumented with the policy variable. Individual-level controls: Marital Status (=1 if Married), Gender (=1 if Female), Education Levels (Base Group is illiterate with no formal education), # of Household Residents. Panel A is estimated using Ordinary Least Squares (OLS) with Community, Year, and Community*Year FE. Panel B is estimated using Two-Stage Least Squares (2SLS) with Community, Year, and Community×Year FE. Clustered standard errors at the community level are reported in parenthesis.
***Significant at the 1 percent level.
**Significant at the 5 percent level.
*Significant at the 10 percent level

# Supplementary Tables
# Online Appendix B

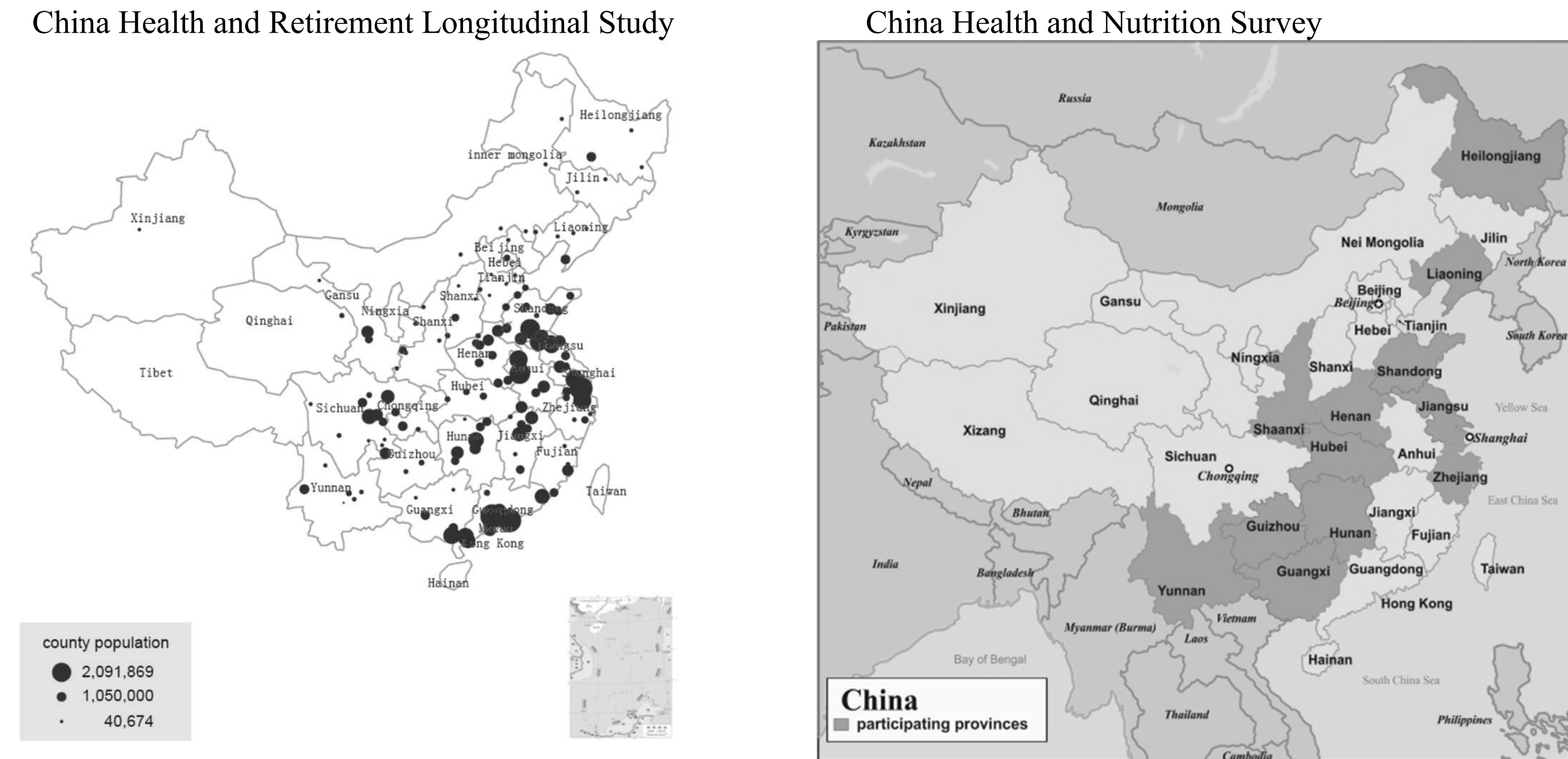

**Fig B1.** Coverage Maps. Source: China Center for Economic Research (2013) and UNC-Carolina Population Center (2015).

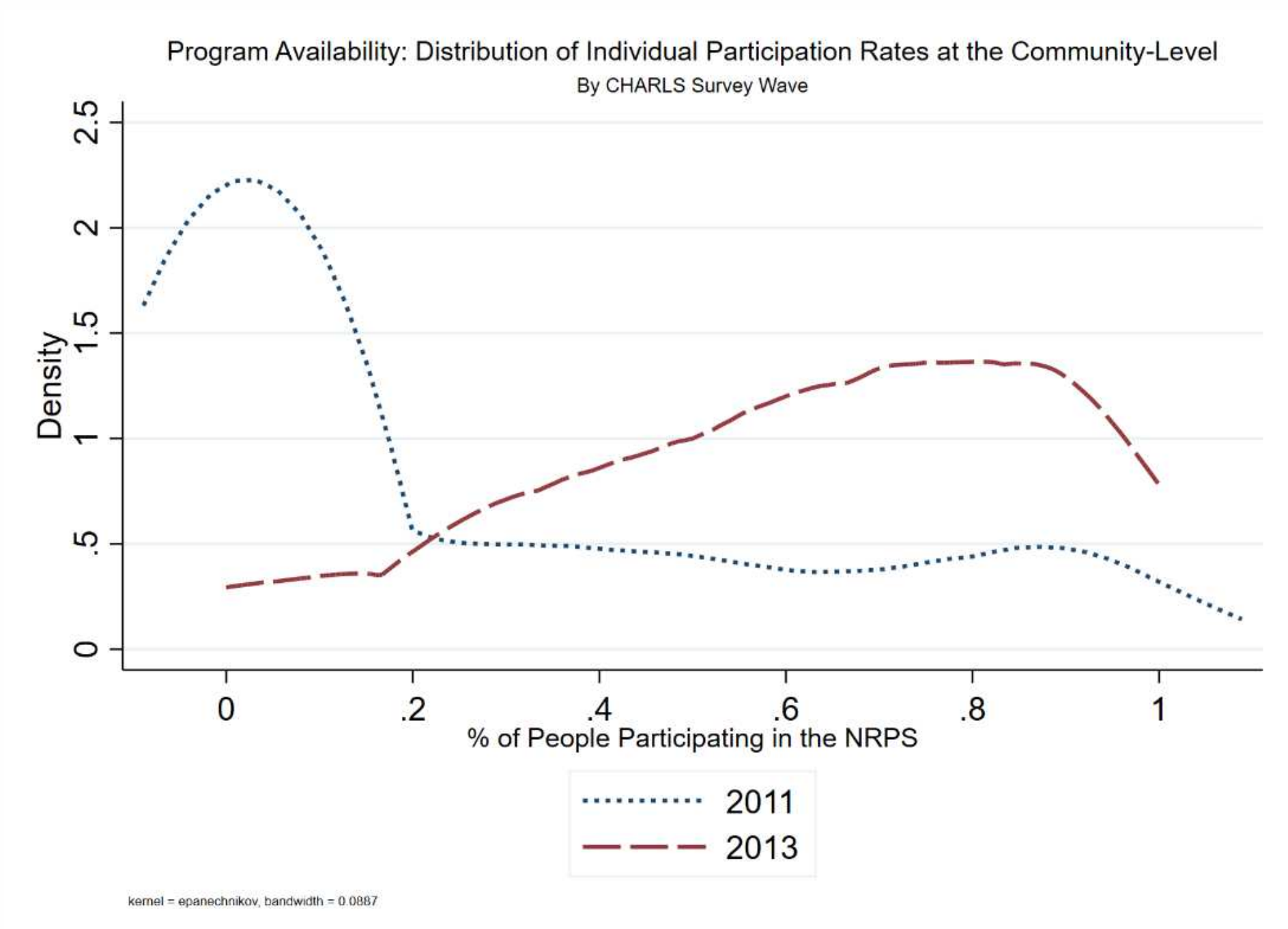

**Fig B2.** Program Availability: Distribution of Individual Participation Rates at the Community Level. Source: CHARLS Survey.

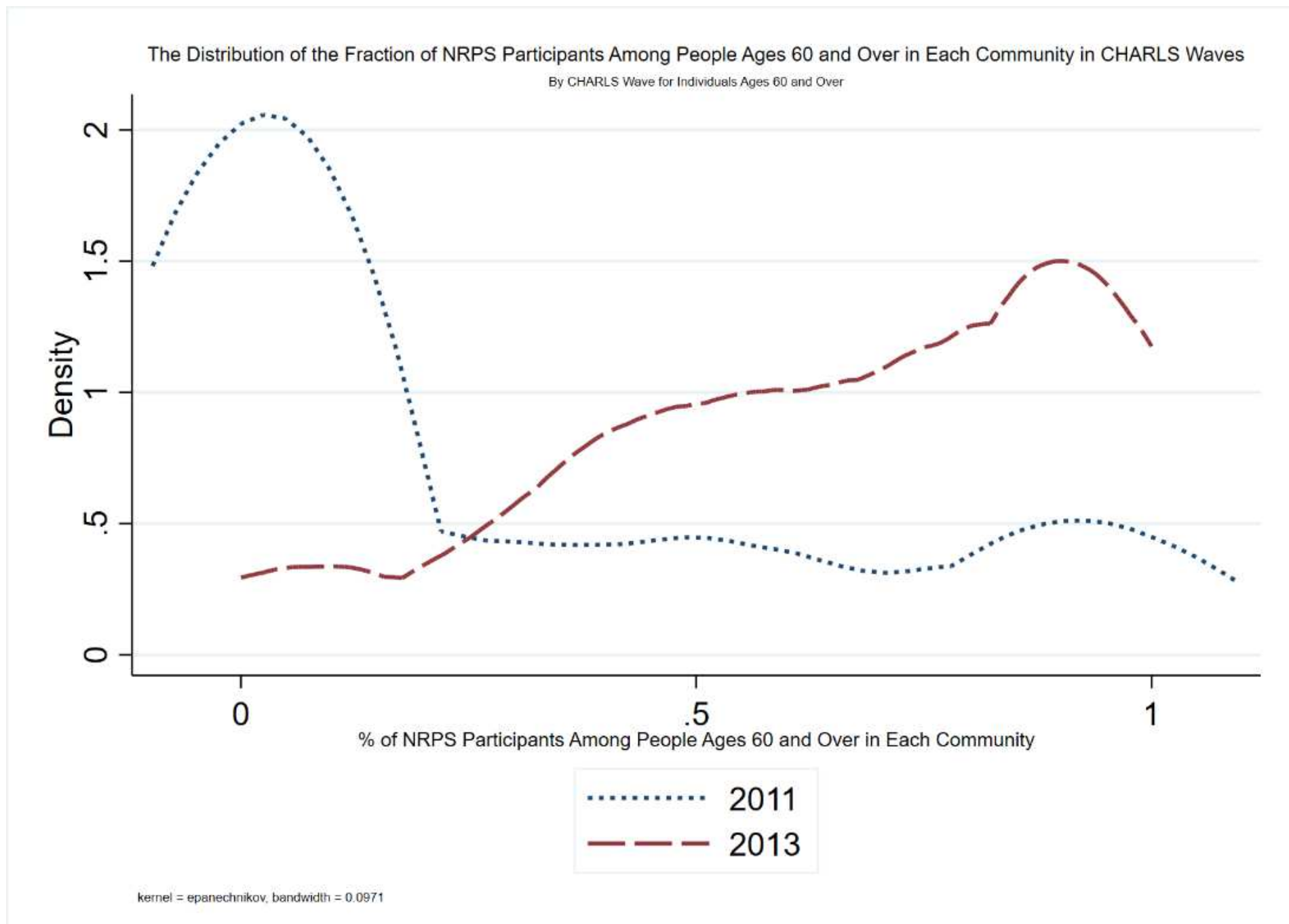

**Fig B3.** Program Availability: The distribution of the fraction of over 60 people in each community who indicate NRPS participation in each CHARLS wave. Source: CHARLS Survey.

**Table B1:** Demographic Characteristics between CHARLS and CHNS Datasets (Age 45 and above and Living in Rural Districts).

| Variables | CHARLS | | | CHNS | | |
|---|---|---|---|---|---|---|
| | # Observations | Mean | Std. Dev. | # Observations | Mean | Std. Dev. |
| | (1) | (2) | (3) | (4) | (5) | (6) |
| Age | 30,590 | 59.27 | 10.01 | 17,130 | 57.65 | 11.61 |
| Han Nationality | 27,280 | 0.922 | 0.268 | 17,086 | 0.847 | 0.360 |
| Female | 30,644 | 0.522 | 0.500 | 17,136 | 0.523 | 0.500 |
| Currently Married | 27,309 | 0.796 | 0.403 | 15,870 | 0.842 | 0.364 |
| At Least Lower Secondary Education | 30,584 | 0.473 | 0.499 | 15,921 | 0.371 | 0.483 |
| # of Children | 27,346 | 2.784 | 1.45 | 15,080 | 2.170 | 1.059 |
| Household Income (Yuan) | 19,701 | 25800.67 | 64097.47 | 15,913 | 27440.65 | 39603.94 |

*Notes*: Yuan reported are in constant 2011 Yuan.

**Table B2:** Component Loadings.

| Cognitive Index | |
|---|---|
| Variable | Loading |
| Immediate Word Recall | 0.595 |
| Delayed Word Recall | 0.588 |
| Serial 7 | 0.414 |
| Self-Reported Memory | 0.137 |
| Knows Current Month (Yes=1) | 0.331 |

**Table B3:** ITT and LATE Estimates on Cognition using Propensity Score for NRPS Participation.

| | Immediate Word Recall[a] | Delay Word Recall[a] | Total Recall[a] | Cognitive Memory Index[b] |
|---|---|---|---|---|
| | (1) | (2) | (3) | (4) |
| Panel A (ITT): | | | | |
| | **NRPS (=1 if Propensity >= Mean + .5 SD** | | | |
| Offered NRPS × Above60[c] | -0.052 | -0.182*** | -0.234** | -0.054 |
| | (0.053) | (0.060) | (0.103) | (0.042) |
| Baseline Mean | 3.792 | 2.862 | 6.678 | 0.000 |
| Controls | Yes | Yes | Yes | Yes |
| R-squared | 0.238 | 0.220 | 0.254 | 0.317 |
| Observations | 18,487 | 18,487 | 18,487 | 18,487 |
| | **NRPS (=1 if Propensity >= Mean + 1 SD** | | | |
| Offered NRPS × Above60[c] | -0.075 | -0.198*** | -0.273*** | |
| | (0.052) | (0.059) | (0.101) | -0.079* (0.041) |
| Baseline Mean | 3.792 | 2.862 | 6.678 | 0.000 |
| Controls | Yes | Yes | Yes | Yes |
| R-squared | 0.230 | 0.216 | 0.247 | 0.310 |
| Observations | 20,309 | 20,309 | 20,309 | 20,309 |
| Panel B (TOT): | | | | |
| | **NRPS (=1 if Propensity >= Mean + .5 SD** | | | |
| PrNRPS × Above60[d] | -0.112 | -0.396*** | -0.508** | -0.118 |
| | (0.116) | (0.133) | (0.226) | (0.091) |
| Baseline Mean | 3.792 | 2.862 | 6.678 | 0.000 |
| Controls | Yes | Yes | Yes | Yes |
| F-Stat (First Stage) | 458.733 | 458.733 | 458.733 | 458.733 |
| Observations | 18,487 | 18,487 | 18,487 | 18,487 |
| | **NRPS (=1 if Propensity >= Mean + 1 SD** | | | |
| PrNRPS × Above60[d] | -0.161 | -0.427*** | -0.588*** | -0.170* |
| | (0.112) | (0.129) | (0.220) | (0.088) |
| Baseline Mean | 3.792 | 2.862 | 6.678 | 0.000 |
| Controls | Yes | Yes | Yes | Yes |
| F-Stat (First Stage) | 348.111 | 348.111 | 348.111 | 348.111 |
| Observations | 20,309 | 20,309 | 20,309 | 20,309 |

*Notes*: The table reports estimates of the DDD estimator for the NRPS treatment effect. (a) Word recall tests: Immediate Recall = [0,10], Delayed Recall = [0,10] and Total Recall = [0,20]. (b) We created the Cognitive Memory Index using principal component analysis, combing measures of short/long term memory, working memory, and orientation. (c) Our DDD coefficient: Policy instrument interacted with an indicator for being over 60 years old. (d) Individual-level participation variable constructed from propensity score is instrumented with the policy instrument. Individual-level controls: Above60 (1= Yes), Marital Status (=1 if Married), Gender (=1 if Female), Education Levels (Base Group is illiterate with no formal education), # of Household Residents. Panel A is estimated using Ordinary Least Squares (OLS) with Community, Year, and Community*Year FE. Panel B is estimated using Two-Stage Least Squares (2SLS) with Community, Year, and Community*Year FE. Clustered standard errors at the community level are reported in parenthesis.

***Significant at the 1 percent level.

**Significant at the 5 percent level.

*Significant at the 10 percent level

**Table B4:** ITT and LATE Estimates on Cognition Omitting Particular Communities.

| | Immediate Word Recall[a] | Delay Word Recall[a] | Total Recall[a] | Cognitive Memory Index[b] |
|---|---|---|---|---|
| | (1) | (2) | (3) | (4) |
| **Panel A (ITT):** | | | | |
| | Sample excluding communities with less than 4 participants | | | |
| Offered NRPS × Above60[c] | -0.093* | -0.194*** | -0.287*** | -0.091** |
| | (0.054) | (0.058) | (0.101) | (0.040) |
| Baseline Mean | 3.792 | 2.862 | 6.678 | 0.000 |
| Controls | Yes | Yes | Yes | Yes |
| R-squared | 0.226 | 0.213 | 0.245 | 0.310 |
| Observations | 19,566 | 19,566 | 19,566 | 19,566 |
| | Sample excluding communities with less than 7 participants | | | |
| Offered NRPS × Above60[c] | -0.100* | -0.198*** | -0.297*** | -0.089** |
| | (0.054) | (0.058) | (0.102) | (0.041) |
| Baseline Mean | 3.792 | 2.862 | 6.678 | 0.000 |
| Controls | Yes | Yes | Yes | Yes |
| R-squared | 0.225 | 0.211 | 0.243 | 0.307 |
| Observations | 19,057 | 19,057 | 19,057 | 19,057 |
| **Panel B (TOT):** | | | | |
| | Sample excluding communities with less than 4 participants | | | |
| NRPS Participation × Above60[d] | -0.228* | -0.472*** | -0.700*** | -0.221** |
| | (0.130) | (0.141) | (0.246) | (0.098) |
| Baseline Mean | 3.792 | 2.862 | 6.678 | 0.000 |
| Controls | Yes | Yes | Yes | Yes |
| F-Stat (First Stage) | 243.0807 | 243.0807 | 243.0807 | 243.0807 |
| Observations | 19,566 | 19,566 | 19,566 | 19,566 |
| | Sample excluding communities with less than 7 participants | | | |
| NRPS Participation × Above60[d] | -0.238* | -0.472*** | -0.711*** | -0.213** |
| | (0.130) | (0.139) | (0.243) | (0.098) |
| Baseline Mean | 3.792 | 2.862 | 6.678 | 0.000 |
| Controls | Yes | Yes | Yes | Yes |
| F-Stat (First Stage) | 260.183 | 260.183 | 260.183 | 260.183 |
| Observations | 19,057 | 19,057 | 19,057 | 19,057 |

*Notes*: The table reports estimates of the DDD estimator for the NRPS treatment effect. (a) Word recall tests: Immediate Recall = [0,10], Delayed Recall = [0,10] and Total Recall = [0,20]. (b) We created the Cognitive Memory Index using principal component analysis, combing short/long-term memory measures, working memory (Serial -7 Test) and orientation (Knowing the Current Month), and self-rated memory. (c) DDD coefficient: NRPS availability interacted with an indicator for being over 60 years old. Individual-level controls: Above60 (1= Yes), Marital Status (=1 if Married), Gender (=1 if Female), Education Levels (Base Group is illiterate with no formal education), # of Household Residents. Panel A is estimated using Ordinary Least Squares (OLS) with Community, Year, and Community×Year FE. Panel B is estimated using Two-Stage Least Squares (2SLS) with Community, Year, and Community×Year FE. Clustered standard errors at the community level are reported in parenthesis.
***Significant at the 1 percent level.
**Significant at the 5 percent level.
*Significant at the 10 percent level

**Table B5:** ITT and LATE Estimates on Direct Measures of Health Varying the Definition of Instrument.

| | Immediate Word Recall[a] | Delay Word Recall[a] | Total Recall[a] | Cognitive Memory Index[b] |
|---|---|---|---|---|
| | (1) | (2) | (3) | (4) |
| Panel A (ITT): | | | | |
| | Offer NRPS (=1 if at least 4 in community participate) | | | |
| Offered NRPS × Above60[c] | -0.070 | -0.198*** | -0.268*** | -0.088** |
| | (0.049) | (0.057) | (0.096) | (0.039) |
| Baseline Mean | 3.792 | 2.862 | 6.678 | 0.000 |
| Controls | Yes | Yes | Yes | Yes |
| R-squared | 0.229 | 0.216 | 0.247 | 0.313 |
| Observations | 21,202 | 21,202 | 21,202 | 21,202 |
| | Offer NRPS (=1 if at least 7 in community participate) | | | |
| Offered NRPS × Above60[c] | -0.061 | -0.178*** | -0.239** | -0.078** |
| | (0.049) | (0.057) | (0.096) | (0.039) |
| Baseline Mean | 3.792 | 2.862 | 6.678 | 0.000 |
| Controls | Yes | Yes | Yes | Yes |
| R-squared | 0.229 | 0.215 | 0.247 | 0.313 |
| Observations | 21,202 | 21,202 | 21,202 | 21,202 |
| Panel B (TOT): | | | | |
| | Offer NRPS (=1 if at least 4 in community participate) | | | |
| NRPS Participation × | -0.162 | -0.458*** | -0.620*** | -0.204** |
| Above60[d] | (0.115) | (0.135) | (0.226) | (0.091) |
| Baseline Mean | 3.792 | 2.862 | 6.678 | 0.000 |
| Controls | Yes | Yes | Yes | Yes |
| F-Stat (First Stage) | 240.613 | 240.613 | 240.613 | 240.613 |
| Observations | 21,202 | 21,202 | 21,202 | 21,202 |
| | Offer NRPS (=1 if at least 7 in community participate) | | | |
| NRPS Participation × | -0.140 | -0.405*** | -0.544** | -0.176** |
| Above60[d] | (0.113) | (0.132) | (0.222) | (0.089) |
| Baseline Mean | 3.792 | 2.862 | 6.678 | 0.000 |
| Controls | Yes | Yes | Yes | Yes |
| F-Stat (First Stage) | 258.336 | 258.336 | 258.336 | 258.336 |
| Observations | 21,202 | 21,202 | 21,202 | 21,202 |

*Notes*: The table reports estimates of the DDD estimator for the NRPS treatment effect. (a) Word recall tests: Immediate Recall = [0,10], Delayed Recall = [0,10] and Total Recall = [0,20]. (b) We created the Cognitive Memory Index using principal component analysis, combing short/long-term memory measures, working memory (Serial -7 Test) and orientation (Knowing the Current Month), and self-rated memory. (c) DDD coefficient: NRPS availability interacted with an indicator for being over 60 years old. Individual level controls: Above60 (1= Yes), Marital Status (=1 if Married), Gender (=1 if Female), Education Levels (Base Group is illiterate with no formal education), # of Household Residents. Panel A is estimated using Ordinary Least Squares (OLS) with Community, Year, and Community×Year FE. Panel B is estimated using Two-Stage Least Squares (2SLS) with Community, Year, and Community×Year FE. Clustered standard errors at the community level are reported in parenthesis.
***Significant at the 1 percent level.
**Significant at the 5 percent level.
*Significant at the 10 percent level.

**Table B6:** ITT and LATE Estimates on Cognition. City-Level Analysis (Heilongjiang).

| | Immediate Word Recall[a] | Delay Word Recall[a] | Total Recall[a] | Cognitive Memory Index[b] |
|---|---|---|---|---|
| | (1) | (2) | (3) | (4) |
| Panel A (ITT): | | | | |
| | CHARLS Data | | | |
| Offered NRPS × Above60[c] | -0.317 | -0.151 | -0.468 | -0.213 |
| | (0.267) | (0.263) | (0.476) | (0.189) |
| Baseline Mean | 3.792 | 2.862 | 6.678 | 0.000 |
| Controls | Yes | Yes | Yes | Yes |
| R-squared | 0.113 | 0.080 | 0.106 | 0.149 |
| Observations | 178 | 178 | 178 | 178 |
| | Admin Data | | | |
| Offered NRPS × Above60[c] | 0.272 | -0.288 | -0.016 | -0.057 |
| | (0.693) | (0.762) | (1.308) | (0.507) |
| Baseline Mean | 3.792 | 2.862 | 6.678 | 0.000 |
| Controls | Yes | Yes | Yes | Yes |
| R-squared | .107 | .079 | .099 | .145 |
| Observations | 178 | 178 | 178 | 178 |
| Panel B (TOT): | | | | |
| | CHARLS Data | | | |
| NRPS Participation × | -1.558 | -0.741 | -2.299 | -1.047 |
| Above60[d] | (1.374) | (1.278) | (2.369) | (0.966) |
| Baseline Mean | 3.792 | 2.862 | 6.678 | 0.000 |
| Controls | Yes | Yes | Yes | Yes |
| F-Stat (First Stage) | 10.075 | 10.075 | 10.075 | 10.075 |
| Observations | 178 | 178 | 178 | 178 |
| | Admin Data | | | |
| NRPS Participation × | -0.557 | 0.590 | 0.033 | 0.117 |
| Above60[d] | (1.407) | (1.595) | (2.682) | (1.044) |
| Baseline Mean | 3.792 | 2.862 | 6.678 | 0.000 |
| Controls | Yes | Yes | Yes | Yes |
| F-Stat (First Stage) | 8.526 | 8.526 | 8.526 | 8.526 |
| Observations | 178 | 178 | 178 | 178 |

*Notes*: The table reports estimates of the DDD estimator for the NRPS treatment effect. (a) Word recall tests: Immediate Recall = [0,10], Delayed Recall = [0,10] and Total Recall = [0,20]. (b) We created the Cognitive Memory Index using principal component analysis, combing short/long-term memory measures, working memory (Serial -7 Test) and orientation (Knowing the Current Month), and self-rated memory. (c) DDD coefficient: NRPS availability interacted with an indicator for being over 60 years old. Individual-level controls: Above60 (1= Yes), Marital Status (=1 if Married), Gender (=1 if Female), Education Levels (Base Group is illiterate with no formal education), # of Household Residents. Panel A is estimated using Ordinary Least Squares (OLS) with Community, Year, and Community×Year FE. Panel B is estimated using Two-Stage Least Squares (2SLS) with Community, Year, and Community×Year FE. Clustered standard errors at the community level are reported in parenthesis.
***Significant at the 1 percent level.
**Significant at the 5 percent level.
*Significant at the 10 percent level.